\documentstyle[aps,version2]{revtex}  

\begin{document}


\begin{title}
\centerline{Correlation between the Josephson coupling energy} 
\centerline{and the condensation energy in bilayer cuprate superconductors} \\

\noindent 
\author{Dominik Munzar$^{1),\,2)}$,  
Christian Bernhard$^{2)}$, 
Todd Holden$^{2)}$,\\
Andrzej Golnik$^{2\,*)}$, 
Josef Huml{\'\i}{\v c}ek$^{1)}$, 
Manuel Cardona$^{2)}$}

\begin{instit}
\noindent 
$^{1)}$ Department of Solid State Physics, 
Faculty of Science, 
Masaryk University, \\
Kotl{\'a}{\v r}sk{\'a} 2, CZ-61137 Brno, Czech Republic \\
$^{2)}$ Max-Planck-Institut f\"ur Festk\"orperforschung, 
Heisenbergstra{\ss}e 1, D-70569 Stuttgart, Germany 
\end{instit} 

\end{title}

\begin{abstract}
We review  
some previous studies concerning the intra-bilayer Josephson plasmons
and present new ellipsometric data  
of the $c$-axis infrared response 
of almost optimally doped Bi$_{2}$Sr$_{2}$CaCu$_{2}$O$_{8}$.  
The $c$-axis conductivity of this compound  
exhibits the same kind of anomalies as 
that of underdoped YBa$_{2}$Cu$_{3}$O$_{7-\delta}$.
We analyze these anomalies in detail 
and show  
that they can be explained within a model 
involving the intra-bilayer Josephson effect and 
variations of the electric field inside the unit cell.  
The Josephson coupling energies of different bilayer compounds 
obtained from the optical data 
are compared   
with the condensation energies 
and it is shown 
that there is a reasonable agreement 
between the values of the two quantities.
We argue 
that the Josephson coupling energy,
as determined 
by the frequency of the intra-bilayer Josephson plasmon, 
represents a reasonable estimate 
of the change of the effective $c$-axis kinetic energy 
upon entering the superconducting state. 
It is further explained 
that this is not the case 
for the estimate based on the use  
of the simplest ``tight-binding'' sum rule. 
We discuss possible interpretations 
of the remarkable agreement between the Josephson coupling energies 
and the condensation energies. 
The most plausible interpretation 
is that the interlayer tunneling of the Cooper pairs 
provides the dominant contribution 
to the condensation energy of the bilayer compounds; 
in other words that 
the condensation energy of these compounds
can be accounted for by the interlayer tunneling theory. 
We suggest an extension of this theory, 
which may also explain the high values of $T_{c}$ 
in the single layer compounds 
Tl$_{2}$Ba$_{2}$CuO$_{6}$ and HgBa$_{2}$CuO$_{4}$, 
and we make several experimentally verifiable predictions. 

\end{abstract}


\section{Introduction}

The interlayer tunneling (ILT) theory \cite{Chakravarty1,PWA1,Kumar} provides  
a simple explanation of the surprisingly high values of $T_{c}$ 
in the cuprate superconductors. 
It is based on the idea \cite{Chakravarty1,PWA1,Kumar,Chakravarty2} 
that the pairing mechanism 
is substantially amplified 
by a decrease of the effective $c$-axis kinetic energy of the electrons 
upon entering the superconducting state. 
A prerequisite for this decrease is the absence  
(or at least a strong suppression) 
of the coherent single particle tunneling between the copper-oxygen planes. 
The onset of the Cooper pair tunneling 
by the Josephson mechanism at $T_{c}$ 
then leads to the decrease of the kinetic energy. 
According to the ILT theory, 
this gain of energy, which can be expressed \cite{PWA2} 
as the negatively taken coupling energy 
of the internal Josephson junctions ($E_{J}$),  
represents the dominant part of the condensation energy of the superconductor
($U_{0}$), 
$$
E_{J}\approx U_{0}\,. 
\eqno (1)
$$
This relation, which should be exact in the limit  
of negligible in-plane contribution to the condensation energy 
and negligible coherent single particle tunneling,  
has been shown to be only moderately violated 
for La$_{2-x}$Sr$_{x}$CuO$_{4}$ \cite{Schutzman1}
but strongly violated 
for Tl$_{2}$Ba$_{2}$CuO$_{6}$ (Tl-2201) \cite{Schutzman1,Moler,Tsvetkov1,Dulic}
and for HgBa$_{2}$CuO$_{4}$ (Hg-1201) \cite{Kirtley}. 
Here we show that Eq.~(1) 
is fulfilled for two compounds that 
have two copper-oxygen planes per unit cell (bilayer compounds): 
YBa$_{2}$Cu$_{3}$O$_{7-\delta}$ (Y123)  
and Bi$_{2}$Sr$_{2}$CaCu$_{2}$O$_{8}$ (Bi-2212). 

So far the relation (1) has been tested only 
for single layer compounds, 
since it was not clear 
what kind of coupling takes place 
for the closely spaced copper-oxygen planes of the bilayer compounds. 
Anderson assumed that even these planes are only weakly coupled
and argued that the bilayer compounds consist 
of two kinds of Josephson junctions: inter-bilayer and intra-bilayer 
\cite{PWA3}.  
The Josephson plasma frequency of the inter-bilayer junction ($\omega_{int}$)
is lower than that of the intra-bilayer junction ($\omega_{bl}$). 
The former determines the $c$-axis penetration depth 
while the latter determines $U_{0}$.  
Van der Marel and Tsvetkov \cite{VdMarel1} 
proposed a phenomenological model 
of the dielectric response of such a superlattice
of inter-bilayer and intra-bilayer Josephson junctions.  
They showed that it exhibits a transverse resonance 
between the two zero crossings corresponding to the two plasmons.  
It has been suggested \cite{VdMarel2} 
that this new excitation (``transverse plasma excitation''), 
which can be visualized as a resonant oscillation 
of the condensate density between the two closely spaced copper-oxygen planes,
has indeed been observed as an additional absorption peak 
which appears at low temperature 
in the spectra of the infrared $c$-axis conductivity 
of underdoped Y123
\cite{HomesPRL,Homes,Schutzman2,Bernhard1}. 
Very recently, this interpretation has been put on a firm basis 
by a detailed analysis of the $c$-axis conductivity data 
for Y123 with different oxygen concentrations 
\cite{Phonan,Gruninger}. 
Note that the observation of the transverse plasma excitation   
also implies the existence of the second (intra-bilayer) Josephson plasmon 
which is a vital ingredient of the ILT theory. 
This opens a possibility to test also for bilayer compounds 
the relationship between 
the Josephson coupling energy and the condensation energy 
predicted by the ILT theory (Eq.~(1)). 

In this paper we review some 
of the previous experimental observations of the transverse plasma excitation (TPE). 
In addition, new data for almost optimally doped 
Bi-2212 with $T_{c}\approx 91{\rm\,K}$ 
are reported and analyzed
(section II). 
In section III, the Josephson coupling energies 
of different bilayer compounds 
obtained from the $c$-axis conductivity data 
of Refs.~\cite{Phonan,Gruninger,Zelezny1,Zelezny2}
and of the present work
are compared with the condensation energies obtained 
from the specific heat data \cite{Loram1,Loram2}. 
Section IV contains a discussion
of the frequently used sum-rules-based estimates 
of the changes of the $c$-axis kinetic energy 
\cite{Chakravarty3,Basov1,VdMarel3}. 
Finally, an extension of the interlayer tunneling 
theory is proposed in section V, 
which allows one to explain the high values of $T_{c}$ 
in the single layer compounds Tl-2201 and Hg-1201.  

\section{Results and discussion}  

Since the early days of the high-temperature superconductivity, 
it was known 
that some of the infrared active $c$-axis phonon modes 
of the cuprate superconductors 
exhibit changes (so-called ``phonon anomalies'')  
in the vicinity of the superconducting transition temperature
(see, e.g.~Refs.~\onlinecite{Wittlin,Litvinchuk1,Kamba}). 
The most pronounced anomalies have been observed
for underdoped Y123
\cite{Homes,Schutzman2,Bernhard1,Phonan}.  

\subsection{YBa$_{2}$Cu$_{3}$O$_{7-\delta}$}

The anomalies are illustrated in Fig.~1, which shows our experimental spectra
of the $c$-axis conductivity of YBa$_{2}$Cu$_{3}$O$_{6.55}$  
with $T_{c}=53{\rm\,K}$ \cite{Phonan}.
Note the following anomalies. 
(A) As the temperature decreases, 
the oxygen bond-bending mode at $320{\rm\,cm^{-1}}$ 
which involves the in-phase vibration 
of the planar oxygens against the Y-ion and the chain ions 
\cite{Humlicek,Kress,Henn1}
loses most of its spectral weight and softens by almost $20{\rm\,cm^{-1}}$; 
(B) at the same time a new broad absorption peak appears in the spectra 
around $410{\rm\,cm^{-1}}$ and 
(C) the spectral weight of the peak corresponding 
to the apical oxygen mode at $560{\rm\,cm^{-1}}$ decreases. 
The additional absorption peak (feature (B)) 
has been sometimes considered to be a new phonon \cite{Homes}. 
On the other hand, 
van der Marel and coworkers suggested \cite{VdMarel2} 
that it corresponds to the TPE 
of their model \cite{VdMarel1}. 
We believe that this interpretation is correct, 
but one has to keep in mind 
that some part of the spectral weight of the peak ($\sim 50\%$) 
indeed comes from the phonons 
(see Ref.~\cite{Bernhard1} for a detailed discussion). 
With increasing doping the additional peak shifts towards higher frequencies  
and it becomes broader and less pronounced 
\cite{Phonan,Bernhard1}. 
Although the anomalies start to develop above $T_{c}$ 
in strongly underdoped compounds, 
the pronounced and steep changes always occur right at $T_{c}$ \cite{Bernhard1}. 
Similar though less spectacular effects  
have also been observed for other values of $\delta$
\cite{HomesPRL,Homes,Schutzman2,Bernhard1,Phonan,Gruninger,Bernhard2}, 
for several other bilayer compounds 
\cite{Litvinchuk2,Basov2,Reedyk,Bernhard3,Boris},  
and for the trilayer compound Tl$_{2}$Ba$_{2}$Ca$_{2}$Cu$_{3}$O$_{10}$ 
\cite{Zetterer}.  
Also shown in Fig.~1 are the fits obtained using 
the model of Ref.~\onlinecite{Phonan}. 
Since we are going to refer to this model rather frequently, 
we summarize its main ideas in the appendix A. 
It can be seen  
that the model is capable of providing a good fit 
of both the normal and the superconducting state data. 
The same values of the  oscillator strengths of the phonons
have been used for the normal and for the superconducting states. 
This means that the changes of the spectral weight 
are accounted for by the model
rather than simply fitted.  

\subsection{Bi$_{2}$Sr$_{2}$CaCu$_{2}$O$_{8}$}

Very recently,  {\v Z}elezn{\'y}
{\it et al.}~\cite{Zelezny1,Zelezny2} have reported 
similar anomalies  
in the infrared $c$-axis conductivity 
of underdoped Bi-2212. 
The spectra exhibit an increase 
of the electronic background around $450{\rm\,cm^{-1}}$ below $T_{c}$ 
accompanied by characteristic phonon anomalies. 
The increase of the background corresponds to the additional absorption peak 
such as observed in the spectra of underdoped Y123.  
The most pronounced phonon anomaly again consists 
in a sizeable decrease of the spectral weight 
of the $355{\rm\,cm^{-1}}$ phonon mode, which has  
---according to the shell model calculations \cite{Prade}---
a similar eigenvector as the $320{\rm\,cm^{-1}}$ 
mode in Y123. 
The doping and temperature dependence of the anomalies 
is close to that observed in Y123. 
These findings of {\v Z}elezn{\'y} {\it et al.}~confirmed that the anomalies
are a common property of the bilayer cuprate compounds.  

Fig.~2 shows our experimental spectra  
of the $c$-axis optical conductivity 
of an almost optimally doped Bi-2212 single crystal 
with $T_{c}=91{\rm\,K}$  
which have been obtained by ellipsometric measurements 
(see Ref.~\onlinecite{Bernhard1} and references therein 
for a description of the technique).
The measurements have been performed 
on a large $ac$-face of size $4\times0.5 {\rm\,mm^{2}}$. 
As shown in Fig.~2 (b) the spectra exhibit the same kind of anomalies 
as underdoped Y123 
\cite{HomesPRL,Homes,Schutzman2,Bernhard1,Phonan}
and underdoped Bi-2212 \cite{Zelezny1,Zelezny2}. 
Below $T_{c}$ the electronic background increases 
in the frequency region around ${550 \rm\,cm^{-1}}$ (feature (B))
and simultaneously some of the infrared-active phonon modes are renormalized; 
in particular, the phonon mode at $355{\rm\,cm^{-1}}$
loses a large part of its spectral weight (feature (A)). 
As compared to the data for underdoped Bi-2212,
the region of the spectral weight increase 
is shifted towards slightly higher frequencies. 
This can be expected because the frequency of the TPE 
is known to increase with increasing hole doping 
(see Ref.~\onlinecite{Phonan} for a discussion). 
The onset temperature of the anomalies coincides with $T_{c}$ 
(cf.~the insets of Fig.~2 (a) and Fig.~2 (b))
which confirms that they are related to superconductivity. 
Note that some signatures of the anomalies 
can also be found in the reflectivity data of Ref.~\onlinecite{Tajima}. 

Figure 3 (a) displays the region near the anomalies on an enlarged scale
and Fig.~3 (b) shows the fits obtained by using 
a slightly modified version \cite{Zelezny2} 
of the model of Ref.~\onlinecite{Phonan}. 
The modification concerns only the description 
of the featureless electronic background. 
First, the inter-bilayer susceptibility is set equal to zero
($\chi_{int}=0$ in notation of Ref.~\cite{Phonan}). 
This is a reasonable approximation considering 
the very low value 
of the unscreeneed inter-bilayer plasma frequency $\omega_{int}$
($\omega_{int}\leq 20{\rm\,cm^{-1}}$, see Ref.~\onlinecite{Shibata2}). 
Second, the regular part of the intra-bilayer susceptibility  
($\chi_{bl}$) 
has been fitted by a combination of a broad Drude term 
and a broad mid-infrared Lorentz oscillator instead 
of the single very broad Lorentzian centered at high frequencies 
which was used previously. 
Furthermore, we have adopted the simple picture of the phonon eigenvectors 
used in the fits of the spectra 
of underdoped Y123
and underdoped Bi-2212, i.e., 
we assume that the $355{\rm\,cm^{-1}}$ mode corresponds to  
the vibrations of the planar oxygens, 
whereas the modes at $300{\rm\,cm^{-1}}$ and $580{\rm\,cm^{-1}}$ 
correspond to vibrations of the inter-bilayer oxygens, 
i.e., the apical oxygens and the oxygens of the BiO layers.  
The values of the fitting parameters 
and other numerical factors used are summarized in Table I. 
Those corresponding to the room-temperature spectrum  
have been obtained by fitting  
the measured frequency dependence 
of the complex dielectric function in the frequency region  
above $300{\rm\,cm^{-1}}$  
(with $\omega_{bl}=0$ and $S_{bl}=0$, 
$S_{bl}$ being the oscillator strength of the mid-infrared Lorentz oscillator). 
Those used in calculating the $100{\rm\,K}$ spectrum 
have also been obtained by fitting the data 
(with $\omega_{bl}=0$ and  $S_{bl}=0$), 
except for the values of $\varepsilon_{\infty}$, 
$S_{P}$, $S_{1}$, $S_{2}$ (oscillator strengths of the phonons),                   
$\omega_{P}$ (frequency of the oxygen-bond bending mode), 
$\omega_{1}$ and $\gamma_{1}$ (parameters of the $300{\rm\,cm^{-1}}$ mode),
which have been fixed at the room-temperature ones.                                                                         
Finally, the values of $\varepsilon_{\infty}$, 
$S_{P}$, $S_{1}$, $S_{2}$,                         
$\omega_{P}$, $\gamma_{P}$,                                 
$\omega_{1}$,$\gamma_{1}$, and     
$\gamma_{bl}$ (broadening parameter of the broad Drude contribution),            
have been fixed at the $100{\rm\,K}$ ones, 
when fitting the low-temperature spectrum.                         
It can be seen that the most pronounced 
features of the experimental spectra, 
in particular the increase of the electronic background 
around $550{\rm\,cm^{-1}}$ 
and the spectral-weight anomaly of the oxygen bond-bending mode, 
are well reproduced. 

Figure 3 (c) shows the frequency dependence of the difference 
${\sigma_{1}(T=10{\rm\,K}<<T_{c})-\sigma_{1}(T=100{\rm\,K}\approx T_{c})}$. 
The value of the difference is positive 
in the frequency region between $420{\rm\,cm^{-1}}$ and $580{\rm\,cm^{-1}}$, 
whereas it is negative both for lower frequencies 
(in the region between $340{\rm\,cm^{-1}}$ and $420{\rm\,cm^{-1}}$) 
and for higher frequencies 
(in the region between $580{\rm\,cm^{-1}}$ and $650{\rm\,cm^{-1}}$). 
The positive values in the first region are caused mainly  
by the increase of the electronic background below $T_{c}$ 
(i.e., the additional peak 
due to the TPE). 
The negative values in the second region correspond 
to the spectral weight anomaly of the oxygen bond-bending mode.  
The negative values in the third region are mainly due 
to a slight shift of the apical-oxygen mode at $580{\rm\,cm^{-1}}$ 
towards lower frequencies. 
The agreement between the difference of the measured data (full line)
and the difference of the fitted spectra (dashed line) 
is excellent. 
Note, that this agreement has not been achieved 
by changing the values of the parameters of the phonon modes. 
The slight shift of the $580{\rm\,cm^{-1}}$ mode accounts 
only for a part of the spectral weight increase around $550{\rm\,cm^{-1}}$. 

Finally, the frequency dependence of the quantity $N_{n}-N_{s}$, where 
${N_{n}(\omega)=N(T=100{\rm\,K}\approx T_{c},\omega)}$,   
${N_{s}(\omega)=N(T=10{\rm\,K}<<T_{c},\omega)}$ and  
$$
N(T,\omega)=
\int_{0^{+}}^{\omega}\,d\omega'
\sigma_{1}(T,\omega')\,, 
\eqno (2) 
$$ 
is displayed in Fig.~3 (d). We take our low-frequency cutoff 
of $70{\rm\,cm^{-1}}$ 
as the lower limit of the integration in Eq.~(2), 
but it is rather unlikely  
that below this frequency there is any considerable difference  
between the normal state and the superconducting state data. 
For a conventional superconductor, the value of $N_{n}-N_{s}$ 
increases with increasing frequency and it approaches $\rho_{s}$, 
the spectral weight of the $\delta$-peak at $\omega=0$,  
within a range of $\sim 6\Delta$, $\Delta$ 
being the superconducting gap \cite{Tinkham}. 
It has been shown by Basov et al.~\cite{Basov1}
that for several high-temperature superconductors  
the value of $N_{n}-N_{s}$ also increases with increasing frequency 
but it saturates at a value 
of only about one half of $\rho_{s}$ for $\omega\sim 10\Delta$. 
Our result for Bi-2212
exhibits an even more suprising tendency: 
$N_{n}-N_{s}$ stays approximately 
constant up to $\sim 350{\rm\,cm^{-1}}$, it changes considerably 
in the frequency region between $350{\rm\,cm^{-1}}$ and $620{\rm\,cm^{-1}}$ 
and it seems to saturate above $\sim 650{\rm\,cm^{-1}}$ 
at a negative value of approximately $-300\,\Omega{\rm^{-1}\,cm^{-2}}$. 
The corresponding value of the ratio $|N_{n}-N_{s}|/\rho_{s}$ 
is larger than 25.  
The fact that the value of $N_{n}-N_{s}$ above $\sim 550{\rm\,cm^{-1}}$ 
is negative 
signals an increase of the low frequency spectral weight below $T_{c}$ 
which is caused by the formation of the peak due to the TPE. 
Note that such an increase cannot be easily explained by any conventional theory. 
We shall come back to this interesting issue in section IV B. 

\section{Josephson coupling energies and condensation energies} 

In Table II we summarize the values 
of the intra-bilayer plasma frequency ($\omega_{bl}$),
the corresponding Josephson coupling energy per unit cell ($E_{J}$), 
and the condensation energy per unit cell ($U_{0}$) 
for Y123 
and Bi-2212. 
The values of $E_{J}$ have been calculated 
using the formula \cite{EJ}  
$$ 
E_{J}={{\hbar}^{2} \varepsilon_{0} a^{2}\over 4 e^{2} d_{bl}}
                    \omega_{bl}^{2}
\,\,\,\,{\rm or\,equivalently}\,\,\,\,
E_{J}[{\rm\,meV}]={C\over d_{bl}[{\rm \AA}]}\left(\omega_{bl}[{\rm cm^{-1}}]\right)^{2}\,. 
\eqno (3)
$$
Here $a$ is the in-plane lattice constant,   
$d_{bl}$ is the distance between the closely spaced copper-oxygen planes, and  
$C=3.1\times 10^{-7}$. 
We neglect the contribution of the inter-bilayer plasmon to $E_{J}$ 
which is by at least one order 
of magnitude lower than that of the intra-bilayer one.  
Note that Eq.~(3) does not contain any adjustable parameters.   
The values of the condensation energies have been obtained 
from the specific heat data  
in Ref.~\onlinecite{Loram1} 
(Y123) 
and in Ref.~\onlinecite{Loram2} (Bi-2212).  
We estimate the error bars 
of the values of $\omega_{bl}$ 
to be about $20-30\%$. 
These error bars arise mainly from the uncertainty
in the description of the electronic background \cite{background}. 
The corresponding uncertainties 
of the values of $E_{J}$ are about $50\%$. 
Taking this into account, the agreement between 
the Josephson coupling energies obtained from the optical data 
and the condensation energies obtained from the specific heat data 
is reasonably good: \\
(i) The values of the two quantities 
are of the same order of magnitude.\\
(ii) Both quantities exhibit a rather similar dependence on doping. 
It is even possible to understand,
why the values of the ratio $E_{J}/U_{0}$ are higher 
for the strongly underdoped samples than for the less underdoped ones 
or the optimally doped ones.
It is namely likely that for strongly underdoped samples
some fluctuation effects set in 
well above the macroscopic transition temperature $T_{c}$. 
This suggestion is consistent with the finding 
that the phonon anomalies start to occur well above $T_{c}$ 
in strongly underdoped samples \cite{Bernhard1,Phonan,aboveTc}.  
The contribution of the fluctuation effects to the condensation energy 
is not likely to be contained in the values of $U_{0}$ 
presented in Table II, 
since they have been obtained 
by an analysis of the specific heat data 
where only the changes occuring below $T_{c}$ 
are properly taken into account.
On the other hand, the values of $E_{J}$ do contain 
the contribution of these effects,
since they are determined by the low-temperature values of $\omega_{bl}$ 
(i.e., in a way 
which does not require 
any assumptions concerning the onset of superconductivity).  
As a result the values of $U_{0}$ for strongly underdoped samples 
can be expected to be smaller 
than the values of $E_{J}$.\\ 
(iii) The values of $E_{J}$ for Y123 
are systematically higher than those for Bi-2212, 
in agreement with the trend in the condensation energies. 
It may be argued that the difference in the condensation energies 
is related to the presence of metallic chains 
in Y123. However, the chain condensation 
cannot fully account for the difference, since Ca substituted 
samples with broken chains also have a significantly higher condensation 
energy than Bi-2212 \cite{Loram1}.
 
This agreement means that the values of the condensation energies 
of the bilayer compounds can be explained using the ILT theory. 
Further implications will be discussed in Section V.
It certainly remains an open question 
why different compounds 
with almost identical values of $T_{c}$ 
have rather different values of the condensation energy. 

\section{Sum rules} 

There is another approach to estimate 
the changes of the $c$-axis kinetic energy, 
pioneered by Chakravarty, Kee and Abrahams \cite{Chakravarty2,Chakravarty3}, 
which is based 
on the use of the optical sum rule. 
For the bilayer compounds, 
this approach yields  
an opposite result, namely that the value of the condensation 
energy cannot be accounted for 
by the change of the $c$-axis kinetic energy.
In the first part of this section  
we clarify several conceptual points
which have made the discussions of the kinetic energy changes 
somewhat confusing (subsection A). 
Next we discuss some specific properties 
of the bilayer compounds (subsection B).
In particular, we show that  
for these compounds the change of the $c$-axis kinetic energy 
cannot be estimated using the conventional 
``tight-binding'' sum rule of Refs.~\cite{Chakravarty2,Chakravarty3}.   

\subsection{Conceptual issues} 

For simplicity, we focus in our considerations 
first on the single layer compounds. 
In the present context, there are 
at least three different quantities denoted as $c$-axis kinetic energy.  
First, the ``true'' $c$-axis kinetic energy, 
$K_{c}=\sum_{i} p_{i,c}^{2}/2m$.  
Second, the ``tight-binding'' $c$-axis kinetic energy $H_{c}$,  
$$
H_{c}=-\sum_{l,i,j,s}\,t_{\perp}(l,i,j)c_{j,l+1,s}^{+}c_{i,l,s}\,\,+H.c.\,, 
\eqno (4) 
$$ 
where l is the layer index, 
$i$ and $j$ refer to the sites of the two-dimensional layers, 
$s$ refers to spin, $t_{\perp}$ is 
the interlayer hopping matrix element, 
and $H.c.$ means the Hermitian conjugate operator. 
Third, the effective low-energy $c$-axis 
kinetic energy of the ILT theory $H_{J}$, 
$$
H_{J}=-\sum_{l,{\bf k}}\,T_{J}({\bf k})c_{{\bf k},l+1,\uparrow}^{+}
                                 c_{-{\bf k},l+1,\downarrow}^{+}
                                 c_{-{\bf k},l,\downarrow}
                                 c_{{\bf k},l,\uparrow}
\,\,+H.c.\,, 
\eqno (5)
$$
i.e., the Josephson coupling energy. 
The experimental data are discussed in terms of at least two different 
sum rules. First, the ``general'' sum rule (for derivation see, e.g, 
Ref.~\cite{Varenna}),  
$$
\int_{0}^{\infty}\sigma_{1}(T,\omega)\,d\omega={\pi ne^{2}\over 2m}\,, 
\eqno (6)
$$ 
which is fulfilled for any real system
and second, the ``tight-binding'' sum rule valid only for model Hamiltonians 
whose single-particle part is of the tight-binding form 
(for derivation see, e.g., 
Refs.~\onlinecite{Scalapino,Chakravarty2} and references therein),  
$$ 
\int_{0}^{\infty}\sigma_{1}(T, \omega)\,d\omega=
-{\pi e^{2} d \over {2 \hbar}^{2}a^{2}} \langle H_{c}\rangle (T)\,.
\eqno (7)
$$
In Eq.~(6), $n$ is the total electron density,   
in Eq.~(7) $a$ is the in plane lattice constant, 
$d$ is the lattice constant along the $c$-axis, 
and $H_{c}$ is the tight-binding kinetic energy (Eq.~(4)) per unit cell
($\langle H_{c}\rangle $ represents an average of this quantity, 
$T$ is the temperature). 

Let us discuss  
possible changes of the three kinetic energies 
upon entering the superconducting state and 
the relations between them and the sum rules. \\
(i) There is no obvious relation between $\langle K_{c} \rangle$ 
and the sum rules. 
Upon entering the superconducting state,  
the total energy of the superconductor decreases. 
This implies---by virtue of the virial theorem 
(${\langle K \rangle =-\langle V \rangle /2}$, 
where $K$ and $V$ are the kinetic and the potential energy, 
respectively; see, e.g., Ref.~\cite{virialtheorem})---that 
its total kinetic energy increases. 
It is possible to assume that even $\langle K_{c}\rangle $ increases 
at the superconducting transition.\\
(ii) It is very likely that the degrees 
of freedom essential for superconductivity  
are contained in some model single band Hamiltonian $H_{1b}$ 
whose (tight-binding like) single particle part 
is derived by a downfolding process 
in which all of the higher energy bands are integrated out \cite{Andersen} 
and whose interaction part describes the most pronounced correlation effects. 
In our opinion, there are no general reasons 
why the average   
of its $c$-axis kinetic energy term $H_{c}$ (Eq.~(4)) 
should decrease or increase below $T_{c}$. 
The arguments invoking a frustration of the $c$-axis kinetic energy 
in the normal state and a ``deconfinement'' of bosonic pairs below $T_{c}$
(See, e.g., Ref.~\cite{PWA1}, Chapter 2, Dogma VI and 
Ref.~\cite{Kumar}) are usually based on asymptotic properties 
of the quasiparticle propagators, 
i.e., they concern rather the asymptotic behaviour for $\omega\rightarrow 0$ 
than the actual changes of $\langle H_{c} \rangle$. 
Possibly these  arguments could be formulated more precisely in terms  
of yet another effective kinetic energy $H_{c}'$ 
characterising the $c$-axis electrodynamics 
in a frequency scale intermediate between 
that of $H_{c}$ (${\sim 10 000 \rm\,cm^{-1}}$) and 
the frequency scale of $H_{J}$ of the order of $100 {\rm\,cm^{-1}}$ 
\cite{scaleofHj} (see Fig.~4).  
 
It follows from Eq.~(7) that the possible change of $\langle H_{c} \rangle$
at the superconducting transition 
is related to the change of the spectral weight as follows: 
$$ 
\left( 
\langle H_{c}\rangle _{s}(T<<T_{c})-
\langle H_{c}\rangle _{n}(T\approx T_{c})
\right)
[{\rm\,meV}]
=
$$
$$
-{4 C\over d[{\rm \AA}]}{120\over \pi}
\left( 
\int_{0}^{\Omega_{c}}\sigma_{1}(T<<T_{c}, \omega)
\,d\omega
-\int_{0}^{\Omega_{c}}\sigma_{1}(T\approx T_{c}, \omega)
\,d\omega 
\right)\,[\Omega^{-1}{\rm\,cm^{-2}}]\,,  
\eqno (8)
$$ 
where $\Omega_{c}$ is a cutoff frequency required to exhaust the sum rule (7). 
Note that in a model involving strong correlations, the value of $\Omega_{c}$ 
may be much higher than the bandwidth 
of the corresponding noninteracting model.
In addition, this value  
(and also the changes of the kinetic energy themselves)
may depend on the kind of model which is used, 
i.e., on the interaction part of the model Hamiltonian $H_{1b}$. \\
(iii) We emphasize that it is  
the effective $c$-axis kinetic energy $H_{J}$ of the ILT theory (Eq.~(5)) 
whose changes at the superconducting transition 
were predicted to be responsible 
for the high values of $T_{c}$ in Ref.~\onlinecite{Chakravarty1}.
The only ``sum rule'', which can be associated 
with $H_{J}$, is Eq.~(3) with $d_{bl}$ substituted by $d$ 
(note that $E_{J}=-\langle H_{J} \rangle$). 
 
What is the experimental status? 
The analysis of Basov and coworkers \cite{Basov1} demonstrates 
that at least for the underdoped high-$T_{c}$ superconductors, 
the low-frequency spectral weight 
$$
\alpha(T,\omega)=
\int_{0}^{\omega}\sigma_{1}(T, \omega')\,d\omega'=
\rho_{s}(T)+N(T,\omega)\, 
\eqno (9) 
$$
increases considerably upon entering the superconducting state 
for cutoff frequencies $\omega$ ranging up to at least $1200{\rm\,cm^{-1}}$. 
More precisely, as one increases the value of $\omega$,  
the value of the difference 
$\alpha(T<<T_{c},\omega)-\alpha(T\approx T_{c},\omega)$
saturates at $\omega\approx {500 \rm\,cm^{-1}}$,  
reaching typically $50\%$ of $\rho_{s}$. 
The only way to reconcile this interesting finding with  
the general sum rule (6) 
consists in assuming \cite{Basov1} that a substantial part 
of $\alpha(T<<T_{c},1200{\rm\,cm^{-1}})$ 
is collected from frequencies exceeding ${1200 \rm\,cm^{-1}}$. 
In other words, 
since the total spectral weight has to be conserved,   
the increase of $\alpha(T,{1200\rm\,cm^{-1}})$
below $T_{c}$ is compensated by a decrease of spectral weight 
around some frequency $\omega^{*}$ higher than ${1200 \rm\,cm^{-1}}$.  
A sketch of the situation is shown in Fig.~4. 

What do we learn from these results about the changes
of the tight-binding kinetic energy $\langle H_{c} \rangle$? 
Let us denote by $\omega_{c}$ the interband cutoff, 
i.e., the upper limit of the frequency interval 
for which the response of the superconductor 
can be properly described using a certain effective Hamiltonian $H_{1b}$
(e.g., the Hubbard Hamiltonian). 
Let us further assume, for the sake of simplicity, 
that $\omega_{c}\geq \Omega_{c}$ of Eq.~(8) \cite{cutoffs}. 
There are two different possibilities, 
which are both compatible with the infrared data: 
(a) $\omega^{*}>\omega_{c}$ and (b) $\omega^{*}<\omega_{c}$ (see Fig.~4).  
In case (a), it follows from Eqs.~(6) and (8)
that $\langle H_{c}\rangle $ indeed decreases below $T_{c}$. 
In order to establish this decrease experimentally, however, 
one would have to estimate the total low-frequency spectral weight 
above and below $T_{c}$, 
i.e., to integrate $\sigma_{1}(\omega)$ up to $\omega_{c}$. 
We believe that the case (b) is more likely to happen. 
The increase of $\alpha(T,{1200\rm\,cm^{-1}})$ below $T_{c}$
and the corresponding decrease of the spectral weight around $\omega^{*}$ 
then both can be described by using $H_{1b}$.  
It follows from Eqs.~(6) and (8) that $\langle H_{c} \rangle$ 
does not change at $T_{c}$.
To conclude, 
it is only the tight-binding $c$-axis kinetic energy $H_{c}$ (Eq.~(4))
of an effective single band Hamiltonian, 
whose change at the superconducting transition 
can be obtained using Eq.~(8). 
The results of Basov et al.~\cite{Basov1} nicely 
reveal the unconventional properties 
of the high-temperature superconductors,  
in particular the extremely large frequency scale involved,  
but they cannot be used to yield reliable estimates of this  
change because of the uncertainties 
concerning the cutoff frequencies $\Omega_{c}$ and $\omega_{c}$. 
In addition, there are no ${\it a priori}$ reasons, 
why $\langle H_{c} \rangle$ should change upon entering 
the superconducting state. 

Another example of a model $c$-axis kinetic energy 
is represented by the effective low-energy $c$-axis kinetic energy $H_{J}$ (Eq.~(5))
of  Ref.~\onlinecite{Chakravarty1}.  
It acquires a nonzero (negative) value, 
$-E_{J}$ (Eq.~(3)), 
only in the superconducting state and  
it was predicted to be responsible for the high values of $T_{c}$. 
It is certainly a crude approximation to identify $E_{J}$ 
with the $c$-axis contribution to the condensation energy,
as we did it in the previous section,  
since possible ``countereffects'' \cite{countereffects} are not included. 
They are to some extent included 
when estimating the changes of the $c$-axis kinetic energy 
using Eq.~(8) with the upper limit
of the integrals well below $\omega^{*}$ (instead of $\omega_{c}$), i.e., 
$$ 
\Delta E_{kin,c}[{\rm meV}]=
-{4 C\over d[{\rm \AA}]}{120\over \pi}
\left(
\alpha(T<<T_{c}, \omega<\omega^{*})-
\alpha(T\approx T_{c}, \omega<\omega^{*})
\right)
\,[\Omega^{-1}{\rm\,cm^{-2}}]\,,                      
\eqno (10)  
$$
This is an appealing possibility, 
but the physical meaning of the result  
is not completely clear 
(in contrast to the change 
of the well defined quantity $\langle H_{c} \rangle$). 
It may be perhaps viewed as an estimate of the change 
of some kinetic energy $H_{c}'$ 
characterising the $c$-axis electrodynamics 
in an intermediate frequency scale (see our discussion above
and Fig.~4). 
Note finally that the fact 
that $\Delta E_{kin,c}$ includes the countereffects 
does not necessarily imply
that $|\Delta E_{kin,c}|<E_{J}$ 
because of the additional factor of 4 
entering the sum-rules-based formulas 
\cite{factor4}.
On the contrary,  
for moderate countereffects, $|\Delta E_{kin,c}|>E_{J}$.  
For underdoped La$_{2-x}$Sr$_{x}$CuO$_{4}$, e.g., 
$|\Delta E_{kin,c}|$ is by a factor of $\sim 2$ larger 
than $E_{J}$\cite{factor4}. 

\subsection{Some specific properties of the bilayer compounds} 

In Fig.~3 (d), we have encountered a rather unusual situation. 
First, the value of the quantity $N(T,690{\rm\,cm^{-1}})$ 
defined by Eq.~(2)
increases with decreasing temperature. 
A phenomenon not observed 
for any of the single-layer compounds
(see, e.g., Ref.~\onlinecite{Basov1}). 
Second, the increase of the integrated spectral weight  
$\alpha(T,690{\rm\,cm^{-1}})$ defined by Eq.~(9) 
upon entering the superconducting state 
is lower by a factor of $\sim 30$  
than the value required to yield 
a kinetic energy change comparable to $U_{0}$ (when inserted into Eq.~(10)). 
On the other hand, using simply the formula (3)
for the Josephson coupling energy, 
we have obtained a value fairly close to $U_{0}$. 
Here we propose a qualitative explanation of these paradoxes.  

Let us consider the superlattice of intra-bilayer and inter-bilayer 
Josephson junctions. 
The low frequency spectral weight 
$\alpha(T<<T_{c},\omega>\omega_{p})$, 
where $\omega_{p}$ is the frequency of the TPE (Eq.~(A.2)),  
contains both the contribution of the $\delta$-peak at $\omega=0$, 
$S_{\delta}$ (Eq.~(A.3)), 
and the contribution of the TPE at $\omega=\omega_{p}$, $S_{pl}$ (Eq.~(A.4)). 
The appearance of the second one below $T_{c}$ 
explains the first paradox. 
It is easy to show that 
$$
\alpha(T<<T_{c},\omega>\omega_{p})={\pi\over 2}\varepsilon_{0}
{d_{bl}\omega_{bl}^{2}+d_{int}\omega_{int}^{2}\over d_{bl}+d_{int}}
\approx {\pi\over 2} \varepsilon_{0} 
{d_{bl}\omega_{bl}^{2}\over d_{bl}+d_{int}}\,, 
\eqno (11)
$$ 
where $d_{int}$ is the distance between the bilayers. 
The Josephson coupling energy is approximately given by Eq.~(3) 
(neglecting the contribution of the inter-bilayer Josephson junction), 
whereas the change of the $c$-axis kinetic energy estimated using Eqs.~(10) and (11)
is given as follows: 
$$ 
\Delta E_{kin,c}[{\rm meV}]=
-{4 C\over (d_{bl}+d_{int})[{\rm \AA}]}
{d_{bl} \left( \omega_{bl}[\rm cm^{-1}]\right )^{2}\over d_{bl}+d_{int}}\,.  
\eqno (12) 
$$
It can be seen that 
$$
E_{J}>>|\Delta E_{kin,c}|\,.
\eqno (13)
$$
This is the explanation of the second paradox. 
The effects of the electronic background 
make the discrepancy even more pronounced. 
At the same time, it is obvious  
that within the simple model of the superlattice 
of intra-bilayer and inter-bilayer Josephson junctions, 
it is the Josephson coupling energy $E_{J}$, rather 
than $\Delta E_{kin,c}$, 
which represents the change of the $c$-axis 
kinetic energy upon entering the superconducting state. 
We conclude 
that for the bilayer compounds 
the change of the $c$-axis kinetic energy
can be estimated using Eq.~(3) 
but cannot be estimated using Eq.~(10).  
We refer the reader to appendix B for a more rigorous discussion. 

\section{Possible extension of the ILT scenario 
         for single layer compounds} 

In section III we have shown
that for the bilayer cuprate compounds
the Josephson-coupling energy 
such as estimated  
from the optical data 
can account 
for the condensation energy. 
In section IV B we have shown 
that this result is not invalidated by the fact 
that Eq.~(10) yields   
a much lower estimate 
of the change of the $c$-axis kinetic energy 
upon entering the superconducting state
than Eq.~(3).   
The main reason of the discrepancy 
is that the tight-binding sum rule of Eq.~(7) 
used to derive Eqs.~(8) and (10) is only valid for such single layer compounds, 
for which the distribution of the total electric field is homogeneous.    
The results presented in Section III thus suggest 
that the high-temperature superconductivity in the bilayer compounds 
can be accounted for by the ILT theory. 
We are left with three possible explanations. \\
(a) The agreement between the Josephson-coupling energies 
and the condensation energies reported in section III  
represents a mere coincidence.  \\
(b) The interlayer tunnelling indeed provides 
the dominant contribution to the condensation energy 
of the bilayer compounds.   
Another mechanism is responsible for the high values of $T_{c}$ 
in the single layer compounds Tl-2201 
and Hg-1201. \\                                                    
(c) A modified ILT theory may explain the high-temperature superconductivity 
both in the bilayer compounds and in the single layer compounds.         

In our opinion, it is unlikely that case (a) is realized. 
It would mean either that  
the assignment of the additional absorption peak                             
to the TPE \cite{VdMarel2,Phonan,Gruninger} is wrong                   
or that the Josephson-coupling energy is 
completely (or almost completely) compensated 
by the countereffects \cite{countereffects}. 
There are several important arguments supporting 
the present interpretation of the data.   
Let us mention two of them.
First, the doping dependence. 
The frequency of the maximum of the additional absorption peak 
increases considerably 
with increasing doping. 
This is easy to explain 
within the present scenario 
(the squared frequency of the maximum 
should be proportional to the condensate density) 
but difficult to explain within theories 
where the additional peak is assigned to a pair breaking excitation
\cite{Hastreiter,Ioffe}. 
According to these theories the frequency of the maximum 
should be close to the pair breaking frequency 
observed in other experiments, 
which seems not to be the case for strongly underdoped samples.   
One could argue \cite{Zeyher} 
that the final state interactions may shift 
the peak towards lower frequencies. 
Second, both the frequency and the spectral weight of the additional peak 
are determined by a single parameter,  $\omega_{bl}$ \cite{Y123Tc53}.  
In other theories, at least two parameters are required to fit the data, 
e.g.,  $\Delta$ and $t_{\perp}$.  
Starting from our interpretation of the data, 
large countereffects would represent 
the only possibility to rule out the ILT mechanism 
from the role of the mechanism providing the dominant 
contribution to the condensation energy $U_{0}$. 
Note that the values of the Josephson-coupling energies 
presented in Table II are so high that even sizeable 
countereffects of the order of 80 \% would not invalidate  
our interpretation (see Ref.~\onlinecite{factor4}). 
A possible analysis of the countereffects along the lines 
of appendix B is complicated by the fact 
that the normal-state data for very low temperatures, 
which should be compared to the superconducting-state data, 
rather than the data obtained for temperatures above $T_{c}$,  
are not available.    
If case (b) were realized, there would be two different mechanisms 
causing the high values of $T_{c}$. 
This is certainly not impossible. 
In the remaining part of this section we argue 
that even the possibility (c) should not be excluded. 
We propose an extension of the ILT theory, 
which may explain the high-$T_{c}$ values  
in the single layer compounds Tl-2201 
and Hg-1201. 
For concretness we focus on the Tl-2201 compound, 
for which some optical data are already available 
\cite{Schutzman1,Tsvetkov1,Dulic,Tsvetkov2,Basov1,Katz}.         

According to the ILT theory as formulated, e.g., in Ref.~\onlinecite{PWA1},
the $c$-axis kinetic energy is frustrated in the normal state because 
of the spin-charge separation mechanism.  
The electrons or holes are composite objects 
that cannot escape from the copper-oxygen planes 
(so called ``confinement''). 
For Cooper pairs this confinement is relaxed 
and consequently the $c$-axis kinetic energy decreases
at the superconducting transition.    
To our best knowledge, 
it has been previously always assumed  
that this decrease is related to 
the onset of Josephson tunneling
between the superconducting copper-oxygen planes.   
This is, however, only one possible mechanism for the decrease.  
We suggest that in Tl-2201 
the $c$-axis kinetic energy decreases 
via some delocalization of the Cooper pairs
which involves the apical-oxygen orbitals.   
By the delocalization we mean the fact 
that the superconducting-state wave function acquires 
a larger contribution of the apical-oxygen orbitals 
than the normal-state wave function.  
Quite generally, the $c$-axis kinetic energy decreases 
at the superconducting transition, 
whenever the Cooper pairs are more delocalized 
along the $c$-axis than the single electrons or holes in the normal state. 
Note that our extension of the ILT theory is based 
on a radical form of the ``confinement'' hypothesis.   
Not only the hopping between the copper-oxygen planes 
is assumed to be blocked in the normal state 
but also (at least to some extent)  
the hopping between the orbitals of the copper-oxygen planes 
and the apical-oxygen orbitals. 
The conjecture can be experimentally tested in several ways. 
First we discuss the $c$-axis infrared spectra, 
which may already provide some evidence in favour 
of the proposed scenario.  
Next we briefly mention some other experimentally 
verifiable consequences. 

Recently, some infrared data for Tl-2201 
have been reported \cite{Basov1,Katz}.   
The spectra of the $c$-axis conductivity 
exhibit four peaks corresponding to the $c$-axis infrared phonons  
---instead of five as dictated by symmetry---and 
two other distinct features, 
which have not been discussed previously:\\ 
(i) A pronounced step around $210{\rm\,cm^{-1}}$. 
For convenience, the relevant part of the conductivity spectra 
is shown in the inset of  Fig.~5 (b).
Note that a weak feature at $\sim 230{\rm\,cm^{-1}}$ 
in the grazing incidence reflectivity data of Ref.~\onlinecite{Tsvetkov2}  
may be related just to this step.  
We suggest that the step 
is due to the missing fifth $c$-axis infrared phonon. \\ 
(ii) An increase of the spectral weight 
of the phonon peak at $390{\rm\,cm^{-1}}$ 
upon entering the superconducting state. 
It can be concluded from Fig.~3 of Ref.~\onlinecite{Basov1} 
that the spectral weight 
in the frequency region between $280{\rm\,cm^{-1}}$ and $440{\rm\,cm^{-1}}$ 
slightly (but not negligibly) increases below $T_{c}$.
For convenience, the important part of the data 
is shown in the inset of Fig.~5 (c).  
We estimate that the spectral weight increase represents
3-5\% of the spectral weight of the phonon. \\ 
Below we show that both features (i) and (ii) 
are compatible with the suggested extension of the ILT theory.  

Figure 6 shows the crystal structure of Tl-2201
such as reported in  Ref.~\onlinecite{Shimakawa}. 
The simplest way to explore the possible consequences of our hypothesis 
consists in dividing the structure into regions A and B
(denoted in Fig.~6), 
assuming that they exhibit different properties
(region B being rather insulating and region A being more metallic),   
and applying the model of Ref.~\onlinecite{Phonan}.
In order to be able to apply the model, 
we have to make some assumptions 
concerning the eigenvectors of the phonons. 
Similarly as in case of Bi-2212 
we concentrate on the three phonons located at higher frequencies
(See Fig.~2 of Ref.~\onlinecite{Basov1})                       
presumably corresponding to vibrations of the oxygens: 
the phonon mode at $600{\rm\,cm^{-1}}$, 
the mode at $390{\rm\,cm^{-1}}$, 
and the phonon which shows up as the step at $210{\rm\,cm^{-1}}$.
According to the  shell model calculations of Ref.~\onlinecite{Kulkarni}, 
the highest frequency mode  
involves vibrations of the TlO-layer oxygens,  
the second one mainly vibrations of the apical oxygens, 
and the third one vibrations of the planar oxygens. 
Besides this assignment ({\it assignment A1}), 
we have also considered the possibility 
that the highest frequency mode corresponds to vibrations 
of the apical oxygens (as in many other cuprates) 
and the second one to vibrations 
of the TlO-layer oxygens ({\it assignment A2}).   
In reality the vibrations of the apicals and the TlO-layer oxygens 
are probably strongly coupled. 
Figure 5 shows the results of our simulations. 
Parts (a) and (b) have been obtained using the assignment {\it A1}, 
parts (c) and (d) using the assignment {\it A2}. 
The model in the form presented in Ref.~\onlinecite{Phonan} 
allows us to treat only two phonon modes at the same time
(the ``interface'' one, i.e., the apical-oxygen mode, 
and only one of the other two modes). 
This is the reason why there are two figures for a given assignment, 
each showing only two structures related to the phonons.  
The electronic conductivities of the two regions A and B 
($\sigma_{A}$ and $\sigma_{B}$)
have been modelled, for the sake of simplicity, by broad Drude terms, 
the plasma frequency of the region A, $\Omega_{A}$, being much higher 
than the plasma frequency of the region B. 
The results are shown for two different values of $\Omega_{A}$: 
$\Omega_{A1}$ (dashed line) and $\Omega_{A2}$ (solid line), 
$\Omega_{A1}<\Omega_{A2}$. 
The increase of $\Omega_{A}$ may simulate 
the changes brought about by the superconducting transition: 
the more delocalized the ground state wavefunction, 
the more metallic the region A  
and the higher the value of $\Omega_{A}$. 
For the sake of simplicity, 
we do not consider the changes of $\sigma_{B}$ 
caused by the onset of superconductivity. 
The values of the parameters used are given in Table III. 

It can be seen in Figs.~5 (b) and 5 (d) 
that the lowest frequency mode 
manifests itself as a step rather than as a Lorentzian peak, 
in agreement with the experimental data.
Note that the step is getting less pronounced 
with increasing value of $\Omega_{A}$, i.e., 
with increasing metallicity of the region A. 
This may correlate with an apparent absence of the structure 
in the data for strongly overdoped Tl-2201
(See Fig.~2 of Ref.~\cite{Katz}). 
As seen in Figs.~5 (a) and (c), 
the phonon mode involving vibrations of the TlO-layer 
oxygens exhibits distinct changes with increasing value 
of $\Omega_{A}$, 
in particular its spectral weight increases \cite{Anomaly}.  
This may correspond to the observed spectral weight anomaly 
of the $390{\rm\,cm^{-1}}$ phonon mode
(assuming that the assignment {\it A2} 
is closer to reality than the assignment {\it A1}).  
Finally, the increase of $\Omega_{A}$ 
results in an increase of the electronic background 
around $1000{\rm\,cm^{-1}}$
(concerning the absolute values see Ref.~\onlinecite{absvalues}). 
Such an increase should be observed in future experiments.  
Note that in the present simulations, 
the increase of $\Omega_{A}$ is intentionally relatively high
so that the anomalies are clearly seen. 
Such an increase would correspond 
to a change of the $c$-axis kinetic energy much higher 
(by a factor of $\sim 4$, a rough estimate 
obtained by using Eq.~(B.11)) 
than the condensation energy \cite{Wade}. 
The actual changes of the electronic background to be observed experimentally 
therefore should be considerably smaller 
(e.g., by a factor of 4).  
Also the frequency range of the increase 
may be somewhat different. 
To summarize, our simulations demonstrate  
that the anomalous features (i) 
and (ii) can be explained 
within a model involving the variations of the electric field
inside the unit cell, 
whereas they cannot be easily explained 
in a conventional way.
The important point is that  
if the electric field indeed changes considerably 
inside the unit cell, the changes of the $c$-axis kinetic energy 
upon entering the superconducting state  
cannot be estimated simply 
from the value of the plasma frequency of the superconducting condensate
or by using the modified tight-binding sum rule (10).  

Let us compare the $c$-axis conductivity 
of  Tl-2201 
with that of another single layer compound 
with considerably lower $T_{c}$, Bi$_{2}$Sr$_{2}$CuO$_{6}$ (Bi-2201). 
The spectra for Bi-2201 
exhibit five peaks corresponding to the $c$-axis infrared phonons 
(See Fig.~2 of Ref.~\onlinecite{Tsvetkov2}, 
similar results have been obtained by our group \cite{Bi2201}), 
i.e., there is one phonon peak more than in the spectra for Tl-2201. 
This difference finds a natural explanation within our extension 
of the ILT theory. Bi-2201 has a substantially 
lower condensation energy than Tl-2201 
\cite{Janod,Nyeanchi}. 
This suggests that in Bi-2201
the apical-oxygen orbitals are less accessible for the charge carriers.  
Consequently, the region A is much less metallic than in case
of Tl-2201 and 
the planar-oxygen mode manifests itself 
rather as a peak than as a step feature. 
This is demonstrated in Fig.~5 (d), where the results for 
$\Omega_{A}=1500{\rm\,cm^{-1}}$ (dotted line) and 
$\Omega_{A}=600{\rm\,cm^{-1}}$ (dashed-dotted line)
are shown. 

An observation of the infrared anomalies discussed above 
in more precise future experiments would represent 
a clear but indirect evidence in favour of the suggested 
extension of the ILT theory. 
Direct evidence could be provided by experiments 
probing the occupation of individual atomic orbitals, 
as, e.g., the near x-ray absorption fine structure (NEXAFS) measurements
(see, e.g., Ref.~\onlinecite{Merz}). 
We predict that such experiments will reveal 
a sizeable increase of the number of holes on the apical sites at $T_{c}$ 
accompanied by the corresponding decrease 
of the number of holes in the copper-oxygen planes.   
Note that this prediction does not concern only Tl-2201 
but all the single layer high-$T_{c}$ cuprates.
It may, of course, also be fulfilled for some of the bilayer cuprates. 
The changes of the hole distribution at $T_{c}$ should further result 
in some structural changes, in particular,  
in changes of the distance 
between the apical oxygens and the copper-oxygen planes, 
and in anomalies of some Raman-active phonon modes 
involving vibrations of the apical oxygens. 
Such anomalies may already have been observed \cite{Misochko}. 

\section{Summary} 

The $c$-axis infrared conductivity of optimally doped Bi-2212 
exhibits the same kind of anomalies as that of underdoped Y123.
Below $T_{c}$ the electronic background increases 
in the frequency region around $550{\rm\,cm^{-1}}$ 
and at the same time the oxygen bond-bending mode 
at $355{\rm\,cm^{-1}}$ loses a large part of its spectral weight. 
The anomalies can be explained within a model 
involving the intra-bilayer Josephson effect and 
variations of the electric field inside the unit cell.  
We have compared the Josephson-coupling energies 
of Y123 and Bi-2212 with different oxygen concentrations 
obtained from the optical data 
with the condensation energies obtained from 
the specific heat data and we have found 
that there is a remarkable agreement between the values of the two quantities.
The Josephson coupling energy is shown to represent a reasonable estimate 
of the change of the $c$-axis kinetic energy upon 
entering the superconducting state
and it is also shown  
that the latter quantity cannot be  obtained  
by using the simple ``tight-binding'' sum rule, 
since this sum rule has been derived assuming homogeneous 
distribution of the total electric field within the unit cell. 
The most plausible interpretation of the agreement 
between the Josephson coupling energies 
and the condensation energies 
is that the condensation energy 
of the bilayer compounds can be accounted for
by the interlayer tunneling theory. 
We propose a modification of this theory 
which may also explain the high values of $T_{c}$ 
in Tl-2201 and Hg-1201. 
The main idea is that the $c$-axis kinetic energy 
of these compounds decreases at $T_{c}$ 
via a delocalization of the Cooper pairs 
onto the apical-oxygen orbitals.
We have investigated, 
using a toy model,   
the consequences of this hypothesis 
for the $c$-axis infrared response 
and we have demonstrated that it offers a simple explanation  
of two features observed in the measured $c$-axis conductivity of Tl-2201.  
In addition, we propose that for Tl-2201 an increase 
of the electronic backround around $1000{\rm\,cm^{-1}}$ takes place at $T_{c}$.  
We further predict a sizeable increase at $T_{c}$ of the number of holes 
on the apical sites, related structural changes 
and related anomalies of some Raman-active phonon modes.\\

Note added.
There are some similarities
between our model \cite{Phonan} of the charge dynamics
and the models of Refs.~\cite{Sardar,Halbritter1,Halbritter2,Chelm,Koyama}.
In Ref.~\cite{Halbritter2} it is argued
that the role of the intra-bilayer plasmon
could be played
by an antiphase oscillation between the copper-oxygen planes
and the ordered blocking layer plaques.
After submitting the paper, O.~K.~Andersen
has drawn our attention to a preprint of Pavarini et al.~\cite{Pavarini},
where it is shown
that there is a correlation
between the maximum value of $T_{c}$ for a given compound, $T_{c\,max}$,
and the value of the parameter $r$
which expresses the range of the intra-layer hopping, $t'/t$,
obtained from band structure calculations.
The higher the value of $r$, the higher the value of $T_{c\,max}$.
It remains an open question to what extent 
this correlation is consistent 
with the extension of the ILT theory proposed in Sec.~V.
There we have suggested
that the condensation energy of the single layer compounds
is largely due to a change of the $c$-axis kinetic energy at $T_{c}$
connected with an increase of the number of holes on the apical sites.
The reason for this change is the confinement of the holes 
to the copper oxygen planes in the normal state
and a deconfinement of the Cooper pairs below $T_{c}$.
In case of the single layer systems
the condensation energy---and also $T_{c}$---
thus should be the largest 
(ignoring the material dependence of the orbital energies) 
for the compounds
with the largest value 
of the hopping matrix element $t_{pa}$
between the relevant orbital of a copper-oxygen plane
(predominantly $d_{x^{2}-y^{2}}$)
and the $2p_{z}$ orbital of the neighbouring apical oxygen plane.
The $2p_{z}$ orbital of an apical oxygen
couples to the copper $4s$ orbital
and $t_{pa}$ is thus determined
(a) by the matrix element $t_{sc}$
between the $2p_{z}$ orbital
and the Cu $4s$ orbital
and
(b) by the ratio of the Cu $4s$ to Cu $d_{x^{2}-y^{2}}$ character,
$R$.
The value of $t_{sc}$ decreases \cite{Pavarini}
with increasing distance $d$
between the copper-oxygen plane and the apical-oxygen plane
and it is thus somewhat smaller
for Tl-2201 ($d\approx 2.7\AA$) and Hg-1201 ($d\approx 2.8\AA$)
as compared to LaSrCuO$_{4}$ ($d\approx 2.4\AA$).
On the other hand, the value of $R$ increases
with increasing value of $r$ as $r^{2}$ \cite{Pavarini},
which means that it is much larger
for Tl-2201 ($r\approx 0.33$) 
and Hg-1201 ($r\approx 0.33$)
than for LaSrCuO$_{4}$ ($r\approx 0.17$)\cite{Pavarini}.
For this reason it is possible
that $t_{pa}$ increases with increasing $r$,
i.e., that the correlation reported in Ref.~\cite{Pavarini} 
is consistent with what one would expect 
on the basis of the extended ILT theory.
Obviously, a careful analysis involving 
the details of the Cu $4s$---apical-oxygen $2p_{z}$ bonding 
is required. 

\acknowledgments
We gratefully acknowledge discussions with 
D.~Basov, P.~C{\'a}sek, O.~Dolgov, 
A.~Dubroka, B.~Farid, M.~Gr{\"u}ninger, J.~Halbritter, 
O.~Jepsen, B.~Keimer, D.~van der Marel, 
S.~Tajima, B.~Velick{\'y}, R.~Zeyher,  V.~{\v Z}elezn{\'y},
and M.~L.~Munzarov{\'a}.  
We thank V.~{\v Z}elezn{\'y} for allowing us 
to present here the values 
of the intra-bilayer plasma frequency of underdoped Bi-2212 
prior to publication of Ref.~\onlinecite{Zelezny2}.
We thank O.~K.~Andersen for drawing our attention 
to Ref.~\cite{Pavarini}.
D.~M.~would like to thank B.~Keimer for hospitality 
and support during a stay at MPI Stuttgart,
where a considerable part of the paper 
has been written. T.~H.~thanks the Alexander von Humboldt Foundation 
for support.  

\vskip 0.2 in 

\centerline{\bf Appendix A: The main ideas of the model of Ref.~\onlinecite{Phonan}}
\vskip 0.2 in 

Figure.~7 (a) shows a schematic representation 
of the electronic part of the model of Ref.~\onlinecite{Phonan}. 
The effects related to the ionic degrees of freedom 
will be discussed subsequently. 
Thick horizontal lines correspond to the two-dimensional 
copper-oxygen planes. 
The distance between the closely spaced copper-oxygen planes 
forming a bilayer is denoted by $d_{bl}$, 
the distance between the bilayers is denoted by $d_{int}$. 
If an electric field $E'$ is aplied, the currents $j_{bl}$ 
(intra-bilayer current) and $j_{int}$ (inter-bilayer current)
flow between the planes. 
Since they are not equal, the planes become charged. 
The resulting surface charge density 
which alternates from one plane to the other is denoted by $\kappa$. 
It modifies the average electric field in the intra-bilayer region,  
$E_{bl}=E'+(\kappa/\varepsilon_{0}\varepsilon_{\infty})$, 
whereas it does not change the average electric field 
in the inter-bilayer region, 
$E_{int}=E'$. 
The currents $j_{bl}$ and $j_{int}$ are determined 
by the fields $E_{bl}$ and $E_{int}$, respectively, 
$j_{bl}=\sigma_{bl}E_{bl}$ 
and 
$j_{int}=\sigma_{int}E_{int}$,  
so that a selfconsistent set of equations is obtained. 
Here $\sigma_{bl}$ and $\sigma_{int}$ 
are the intra- and the inter-bilayer conductivities, respectively. 
The equations can be readily solved yielding the following formula
for the macroscopic dielectric function: 
$$ 
\varepsilon(\omega)=(d_{bl}+d_{int})/ 
               \left[{d_{bl}\over \varepsilon_{bl}(\omega)}
                +{d_{int}\over \varepsilon_{int}(\omega)}\right].   
\eqno (A.1)
$$ 
Here 
$\varepsilon_{bl}(\omega)=(i/\varepsilon_{0}\omega)\sigma_{bl}(\omega)$  
and 
$\varepsilon_{int}(\omega)=(i/\varepsilon_{0}\omega)\sigma_{int}(\omega)$. 
For a superlattice of intra-bilayer and inter-bilayer 
Josephson junctions we have 
$\varepsilon_{bl}(\omega)=\varepsilon_{\infty}-
(\omega_{bl}^{2}/\omega^{2})$ 
and 
$\varepsilon_{int}(\omega)=\varepsilon_{\infty}-
(\omega_{int}^{2}/\omega^{2})$.  
The dielectric function of Eq.~(A.1) then exhibits 
two conventional plasma resonances 
at $\omega=\omega_{bl}/\sqrt{\varepsilon_{\infty}}$ 
and at $\omega=\omega_{int}/\sqrt{\varepsilon_{\infty}}$. 
In addition it exhibits a pole at 
$$
\omega=\omega_{p}=
\sqrt{{d_{bl}\omega_{int}^{2}+d_{int}\omega_{bl}^{2}}\over 
{(d_{bl}+d_{int})\varepsilon_{\infty}}},  
\eqno (A.2)
$$ 
corresponding to a resonant oscillation of the condensate density 
between the two closely spaced copper-oxygen planes (transverse plasma excitation). 
The spectral weight of the $\delta$-peak at $\omega=0$ is 
$$
S_{\delta}=(\pi/2)\varepsilon_{0}
{(d_{bl}+d_{int})\omega_{bl}^{2}\omega_{int}^{2}\over 
d_{bl}\omega_{int}^{2}+d_{int}\omega_{bl}^{2}}\,, 
\eqno (A.3)
$$
and the spectral weight of the resonance at $\omega_{p}$ is 
$$
S_{pl}=(\pi/2)\varepsilon_{0}
{d_{bl}d_{int}
(\omega_{bl}^{2}-\omega_{int}^{2})^{2}\over 
(d_{bl}+d_{int})
(d_{bl}\omega_{int}^{2}+d_{int}\omega_{bl}^{2})}\,. 
\eqno (A.4)  
$$ 
Equations (A.1) and (A.2) have been previously derived in another way 
by Van der Marel and Tsvetkov \cite{VdMarel1} 
(see also Ref.~\onlinecite{Gruninger}).  
As far as only the electronic degrees of freedom are concerned, 
the two approaches yield completely equivalent results. 
However, the present approach 
can be more easily extended to incorporate the phonons. 

Assume for a moment 
that the displacements of the ions do not modify 
the electric fields $E_{bl}$ and $E_{int}$.  
The ions located in the intra-bilayer and in the inter-bilayer regions 
experience then the electric fields $E_{bl}$ and $E_{int}$, respectively. 
The ions located in the copper oxygen planes
experience a field $E_{locP}$ which is equal to the average 
of the two fields, 
$E_{locP}=(E_{bl}+E_{int})/2$. 
This can be easily explained using Fig.~7 (a). 
Let us consider, e.g., the ions in the middle copper-oxygen plane. 
They experience the applied field $E'$
and the electric field generated by the charge density $-\kappa$ 
of the upper copper-oxygen plane 
(the contributions of the other planes cancel each other). 
The total electric field acting on these ions is then 
$E_{locP}=E'+(\kappa/2\varepsilon_{0}\varepsilon_{\infty})=
(E_{bl}+E_{int})/2$. 
Even this simplified picture of the electric fields acting on the ions
allows us to explain the spectral weight anomalies.
The spectacular reduction of the spectral weight 
of the $320{\rm\,cm^{-1}}$ phonon mode 
in underdoped Y123 shown in Fig.~1, e.g., 
is due to the fact that in the frequency region around the phonon 
the two fields $E_{bl}$ and $E_{int}$ have opposite signs
(see Ref.~\onlinecite{Phonan} for a discussion). 

In reality, the displacements of the ions modify 
the electric fields $E_{bl}$ and $E_{int}$, 
so that the model equations become slightly more complicated. 
As an example, we show in Fig.~7 (b), 
how the displacement of the planar oxygens of Y123 (O(2) and O(3)) 
from their equlibrium position 
influences the electric field $E_{bl}$. 
We refer the reader to Refs.~\onlinecite{Phonan,PhonanJLTP} 
for further details of the model. 
\vskip 0.2 in 

\centerline{\bf Appendix B: Approximate ``tight-binding'' sum rule 
for bilayer compounds}
\vskip 0.2 in 

In this appendix we put the conclusions of section IV B  
on a more rigorous basis 
and we present a way of estimating 
the contribution of the countereffects \cite{countereffects}. 
We shall discuss a microscopic counterpart 
of the model outlined in Appendix A.  
For this reason, it may be helpful for the reader to follow Fig.~7. 
Let us consider a model defined 
by the following hamiltonian 
$$ 
H=H_{in-plane}+H_{bl}+H_{int}\,,
\eqno (B.1)
$$
where $H_{in-plane}$ contains 
the intra-planar single particle terms and interaction terms,
$H_{bl}$ and $H_{int}$ contain the hopping terms 
between the closely-spaced copper-oxygen planes 
and between the widely spaced ones, respectively, 
$$
H_{bl}=-t_{\perp\,bl}
     \sum_{l=1,3,5, ...\,;i;s}\,c_{i,l+1,s}^{+}c_{i,l,s}\,\,+H.c.\,, 
\eqno (B.2) 
$$ 
$$
H_{int}=-t_{\perp\,int}
     \sum_{l=2,4,6, ... \,;i;s}\,c_{i,l+1,s}^{+}c_{i,l,s}\,\,+H.c.\,. 
\eqno (B.3) 
$$
In the presence of an electric field along the $c$-axis, 
the hamiltonian reads: 
$$ 
H_{A}=H-Na^{2}\sum_{l=1,3,5 ...} 
       \left( d_{bl}j^{p}(l)A_{bl}+
       {e^{2}d_{bl}^{2}\over 2a^{2}\hbar^{2}} k(l)A_{bl}^{2}
       \right) 
       -Na^{2}\sum_{l=2,4,6 ...} 
       \left( d_{int}j^{p}(l)A_{int}+
       {e^{2}d_{int}^{2}\over 2a^{2}\hbar^{2}} k(l)A_{int}^{2}
       \right)\,.   
\eqno (B.4)
$$
Here  
$$
j^{p}(l)={iet_{\perp\,\alpha}\over Na^{2} \hbar}
              \sum_{i,s}
              \left(c_{i,l+1,s}^{+}c_{i,l,s}
                   -c_{i,l,s}^{+} c_{i,l+1,s}
              \right) \,, 
\eqno (B.5) 
$$
where $\alpha=bl$ for $l\in \{1,3,5,...\}$ 
and $\alpha=int$ for $l\in \{2,4,6,...\}$, 
are the $c$-axis paramagnetic current densities.  
Further, 
$$
k(l)=-{t_{\perp\,\alpha}\over N}
              \sum_{i,s}
              \left(c_{i,l+1,s}^{+}c_{i,l,s}
                   +c_{i,l,s}^{+} c_{i,l+1,s}
              \right) 
\eqno (B.6)
$$ 
are the $c$-axis kinetic energies (per unit cell), 
$N$ is the number of lattice sites in one copper-oxygen plane, 
$A_{bl}$ is the vector potential 
in the intra-bilayer region ($A_{bl}=i\omega E_{bl}$),  
$A_{int}$ is the vector potential 
in the inter-bilayer region ($A_{int}=i\omega E_{int}$). 
When studying the optical response of the system, 
selfconsistent values of these two vector potential have to be used
(RPA approximation). 
Using similar manipulations as in Ref.~\onlinecite{Scalapino}, 
we obtain the following relation between 
the averaged total current densities, $j_{\alpha}$, and 
the two electric fields, $E_{\alpha}$, 
$$
j_{\alpha}=\sigma_{\alpha,\beta}E_{\beta}\,.  
\eqno (B.7)
$$  
Here 
$$
j_{\alpha}=\langle j^{p}(l)+
{e^{2}d_{\alpha}^{2}A_{\alpha}\over \hbar^{2} a^{2} d_{\alpha}}k(l) 
\rangle \,,
\eqno (B.8)
$$ 
where $l=1$ for $\alpha=bl$ and $l=2$ for $\alpha=int$.  
Note that all intra-bilayer regions are identical 
and all inter-bilayer regions are identical,
the long-wavelength limit is assumed. 
Finally the four conductivities $\sigma_{\alpha,\beta}$ 
($\alpha,\beta \in \{bl,int\}$) are given as follows: 
$$
\sigma_{\alpha,\beta}(\omega)={\langle k(l)\rangle 
                e^{2}d_{\alpha}\delta_{\alpha,\beta}/(\hbar^{2}a^{2})\,
                +Na^{2}d_{\alpha} \Lambda_{\alpha,\beta}(\omega)
                \over 
                i(\omega+i\delta)}\,.
\eqno (B.9) 
$$ 
Here 
$$
\Lambda_{\alpha,\beta}(\omega)=
{i\over \hbar} 
\sum_{l'} \int_{-\infty}^{\infty}d (t-t') 
\langle [j^{p}(l,t),j^{p}(l',t')] \rangle \Theta(t-t') e^{i\omega (t-t')}\,,
\eqno (B.10) 
$$
where $l=1$ for $\alpha=bl$  
and $l=2$  for $\alpha=int$;
the sum runs over the odd values of $l'$ for $\beta=bl$
and over the even values of $l'$ for $\beta=int$. 
We are not going to discuss the general case here. 
Instead we concentrate on the case where $t_{\perp\,int}<<t_{\perp\,bl}$, 
i.e., the case of negligible inter-bilayer kinetic energy 
$\langle k(2) \rangle$. 
In this case we can neglect $\sigma_{bl,int}$, 
$\sigma_{int,bl}$ and $\sigma_{int,int}$ and we have 
$j_{bl}\approx \sigma_{bl,bl}E_{bl}$, $j_{int}\approx 0$. 
The current-current correlation functions 
of Eq.~(B.10) have the analytic properties required for the validity 
of the sum rules and we obtain  
$$ 
\int_{0}^{\infty}\sigma_{bl}(\omega)\,d\omega=
-{\pi e^{2} d_{bl} \over {2 \hbar}^{2}a^{2}} \langle k(1) \rangle \,, 
\eqno (B.11) 
$$
where $\sigma_{bl}(\omega)\equiv\sigma_{bl,bl}(\omega)$. 
The sum rule (B.11) yields a formula 
for the change of the kinetic energy 
upon entering the superconducting state 
analogous to Eq.~(8) or Eq.~(10)  
($\sigma (\omega)$ is substituted by $\sigma_{bl}(\omega)$ 
and $d$ is substituted by $d_{bl}$). 
For vanishingly small values of the regular part of $\sigma_{bl}$, 
we obtain  
${\Delta \langle k(1) \rangle [{\rm\,meV}]=
-4C\left( \omega_{bl}[{\rm\,cm^{-1}}]\right) ^{2} /d_{bl}[{\rm\,\AA}]}$.  
This is nothing else than Eq.~(3) 
except for the factor of 4 \cite{factor4}. 
In a general case, Eq.~(11)  yields a recipe 
for how to treat the countereffects.

\begin{table}
\caption{Values of the fitting parameters 
and other numerical factors used in fitting 
the measured $c$-axis infrared conductivity 
of optimally doped Bi-2212. 
The plasma frequency and the broadening parameter of the broad Drude peak 
are denoted by $\Omega_{bl}$ and $\gamma_{bl}$, respectively. The meaning 
of the other symbols is the same as in Ref.~\onlinecite{Phonan}. 
The temperatures are given in K, 
the frequencies and the broadening parameters 
in ${\rm\,cm^{-1}}$. 
The values of the numerical factors $\alpha$, $\beta$, $\gamma$ 
\cite{Phonan},                  
$\alpha=2.28$, $\beta=0.64$, $\gamma=1.28$                                      
have been obtained in the same way as in Ref.~\onlinecite{Phonan}                     
(see also Ref.~\onlinecite{albega})                                                   
using the following values of the effective ionic charges:                      
$e^{*}_{Bi}=3$, $e^{*}_{Sr}=2$, $e^{*}_{Ca}=2$, $e^{*}_{Cu}=2$, 
$e^{*}_{O}=-2$. 
The distances between the planes of a bilayer 
and between the neighbouring bilayers are 
$d_{bl}=3.37{\rm\,\AA}$ and $d_{int}=12.03{\rm\,\AA}$, respectively.  
}
\vskip 0.3 in 
\begin{tabular}{ccccccccccccccccc}
{$T$}                        &\multicolumn{1}{c} 
{$\varepsilon_{\infty}$ }    &\multicolumn{1}{c} 
{$\omega_{bl}$ }             &\multicolumn{1}{c} 
{$\Omega_{bl}$ }             &\multicolumn{1}{c} 
{$\gamma_{bl}$ }             &\multicolumn{1}{c} 
{$S_{bl}$ }                  &\multicolumn{1}{c} 
{$\omega_{b}$ }              &\multicolumn{1}{c} 
{$\gamma_{b}$ }              &\multicolumn{1}{c} 
{$S_{P}$ }                   &\multicolumn{1}{c} 
{$S_{1}$ }                   &\multicolumn{1}{c} 
{$S_{2}$ }                   &\multicolumn{1}{c} 
{$\omega_{P}$ }              &\multicolumn{1}{c} 
{$\omega_{1}$ }              &\multicolumn{1}{c} 
{$\omega_{2}$ }              &\multicolumn{1}{c} 
{$\gamma_{P}$ }              &\multicolumn{1}{c} 
{$\gamma_{1}$ }              &\multicolumn{1}{c} 
{$\gamma_{2}$ }    \\        
\tableline 
300 & 3.8 & 0 & 3000 & 1220 & 0 & 0 & 0 &                         
1.05 & 0.24 & 0.50 &                                              
455 & 332 & 639 &  
10 & 36 & 31 \\                                                           
\tableline 
100 & 3.8 & 0 & 3000 & 1300 & 0 & 0 & 0 &
1.05 & 0.24 & 0.50 &
455 & 332 & 638 & 
12 & 35 & 33 \\
\tableline 
10 & 3.8 & 1180 & 2520 & 1300 & 2.04 & 1000 & 900 &
1.05 & 0.24 & 0.50 &
455 & 332 & 636 & 
12 & 34 & 36 \\
\end{tabular}
\vskip 0.1 in 
\end{table} 

\begin{table}
\caption{Values of the intra-bilayer plasma frequency $\omega_{bl}$,    
the Josephson coupling energy $E_{J}$ and the condensation energy $U_{0}$ 
for several bilayer compounds. The values of $\omega_{bl}$ 
are taken from Ref.~\onlinecite{Phonan} (underdoped Y123), 
Ref.~\onlinecite{Gruninger} (optimally doped Y123), 
Ref.~\onlinecite{Zelezny2} (underdoped Bi-2212);  
the value for optimum doped 
Bi-2212 has been obtained as described in the text. 
The values of the condensation energies 
are taken from Ref.~\onlinecite{Loram1} (Y123) 
and Ref.~\onlinecite{Loram2} (Bi-2212).   
}
\vskip 0.3 in 
\begin{tabular}{cccccc}
{compound}                             &\multicolumn{1}{c} 
{$ T_{c} [{\rm K}]$}                   &\multicolumn{1}{c} 
{$\omega_{bl}[{\rm cm^{-1}}]$}         &\multicolumn{1}{c} 
{$E_{J}[{\rm meV}]$ }                  &\multicolumn{1}{c} 
{$U(0) [{\rm meV}]$ }                  &\multicolumn{1}{c} 
{$E_{J}/U(0)$ } \\       
\tableline 
YBa$_{2}$Cu$_{3}$O$_{6.45}$ & 25 & 950 & 0.08 & 0.01 & 8.0  \\
\tableline 
YBa$_{2}$Cu$_{3}$O$_{6.55}$ & 53 & 1200 & 0.13 & 0.05 & 2.6  \\
\tableline 
YBa$_{2}$Cu$_{3}$O$_{6.75}$ & 80 & 1780 & 0.30 & 0.16 & 1.8  \\
\tableline 
YBa$_{2}$Cu$_{3}$O$_{6.93}$ & 91 & 3480 & 1.14 & 0.36 & 3.2  \\
\tableline 
Bi$_{2}$Sr$_{2}$CaCu$_{2}$O$_{8}$  & 60 & 620 & 0.035 & 0.02 & 1.5  \\
\tableline 
Bi$_{2}$Sr$_{2}$CaCu$_{2}$O$_{8}$  & 80 & 970 & 0.085 & 0.06 & 1.4  \\
\tableline 
Bi$_{2}$Sr$_{2}$CaCu$_{2}$O$_{8}$  & 91 & 1180 & 0.13 & 0.13 & 1.0  \\
\end{tabular}
\vskip 0.1 in 
\end{table}
                                                                                       
\begin{table}                                                                  
\caption{Values of the parameters and other numerical factors used 
in modelling the $c$-axis infrared conductivity                                 
of Tl-2201.                                    
The plasma frequency and the broadening parameter 
of $\sigma_{A}$ and $\sigma_{B}$               
are denoted by $\Omega_{A}$ and $\gamma_{A}$
and $\Omega_{B}$ and $\gamma_{B}$, respectively. 
The parameters of the phonons are denoted by
$S_{1}$, $\omega_{1}$, $\gamma_{1}$, 
$S_{2}$, $\omega_{2}$, $\gamma_{2}$,                  
$S_{3}$, $\omega_{3}$, $\gamma_{3}$.                   
The frequencies and the broadening parameters                                   
are given in ${\rm\,cm^{-1}}$.                                                  
The values of the numerical factors $\alpha$, $\beta$, $\gamma$                 
\cite{Phonan},                                          
have been obtained in the same way as in Ref.~\onlinecite{Phonan}                     
(see also Ref.~\onlinecite{albega})                                                   
using the following values of the effective ionic charges:                      
$e^{*}_{Tl}=3$, $e^{*}_{Ba}=2$,  $e^{*}_{Cu}=2$, $e^{*}_{O}=-2$.                
The distances between the apical-oxygen planes                                 
are $d_{A}=5.3{\rm\,\AA}$ and $d_{B}=6.3{\rm\,\AA}$, respectively.
}
\vskip 0.3 in                                                           
\begin{tabular}{ccccccccccccccc}                                      
{$\varepsilon_{\infty}$ }    &\multicolumn{1}{c}                        
{$\Omega_{A1}$ }           &\multicolumn{1}{c}
{$\Omega_{A2}$ }           &\multicolumn{1}{c}
{$\gamma_{A}$ }            &\multicolumn{1}{c}           
{$\Omega_{B}$ }            &\multicolumn{1}{c}
{$\gamma_{B}$ }            &\multicolumn{1}{c}            
{$S_{1}$ }                   &\multicolumn{1}{c}                        
{$S_{2}$ }                   &\multicolumn{1}{c}                        
{$S_{3}$ }                   &\multicolumn{1}{c}                        
{$\omega_{1}$ }             &\multicolumn{1}{c}                        
{$\omega_{2}$ }             &\multicolumn{1}{c}                        
{$\omega_{3}$ }             &\multicolumn{1}{c}                        
{$\gamma_{1}$ }             &\multicolumn{1}{c}                        
{$\gamma_{2}$ }             &\multicolumn{1}{c}                        
{$\gamma_{3}$ }    \\                                                   
\tableline                                                                     
5 & 2500 & 3000 & 2000 & 600 & 2000 &                                          
0.6 & 0.5 & 0.5 &                                                            
300 & 440 & 640 &                                                              
25 & 20 & 25 \\                                                                 
\end{tabular}                                                                  
\vskip 0.1 in                                                                   
\end{table}                                                                    

\figure{
(a) Experimental spectra of the $c$-axis conductivity, 
$\sigma=\sigma_{1}+i\sigma_{2}$, 
of Y123 with $T_{c}=53{\rm\,K}$. 
Experimental data (thick solid lines) and fits 
(thin solid lines) for (b) $T=300{\rm\,K}$ and (c) $T=4{\rm\,K}$. 
The symbols (A), (B), and (C) indicate the most pronounced 
anomalies as discussed in the text. 
A slightly modified version of Fig.~1 from Ref.~\onlinecite{PhonanJLTP}.  
}

\figure{
Experimental spectra of the $c$-axis conductivity 
of almost optimally doped Bi-2212
with $T_{c}=91{\rm\,K}$.  
(a) Data for $T=300{\rm\,K}$ and $T=100{\rm\,K}$. 
(b) Data for $T=100{\rm\,K}$ and $T=10{\rm\,K}$. 
The insets of (a) and (b) display 
the differences ${\sigma_{1}(T=100{\rm\,K})-\sigma_{1}(T=300{\rm\,K})}$
and   
${\sigma_{1}(T=10{\rm\,K})-\sigma_{1}(T=100{\rm\,K})}$, respectively. 
Note that the former and the latter difference corresponds 
to a temperature change of $200{\rm\,K}$ and $90{\rm\,K}$, respectively. 
The symbols (A) and (B) in (b) indicate the most pronounced anomalies 
as discussed in the text. 
}

\figure{
(a) Experimental spectra of the $c$-axis conductivity 
of optimally doped Bi-2212---region 
of the spectra near the anomalies on an enlarged scale. 
(b) Fits of the spectra obtained by using the model of Ref.~\cite{Phonan}
as described in the text. 
(c) Spectra of ${\sigma_{1}(T=100{\rm\,K})-\sigma_{1}(T=10{\rm\,K})}$. 
(d) Spectra of the quantity $N_{n}-N_{s}$ defined in the text. 
} 
 
\figure{
Schematic representation of the spectral-weight changes 
upon entering the superconducting state. 
The increase of spectral weight at low frequencies below $T_{c}$ 
is compensated  by a decrease around some frequency $\omega^{*}$. 
This frequency may be either (a) higher or (b) lower  
than the interband cutoff $\omega_{c}$. 
The horizontal lines below the frequency axis of (b) 
indicate the three frequency scales 
corresponding to the three effective $c$-axis kinetic energies  
discussed in the text.  
}
 
\figure{
Results of model computations of the $c$-axis infrared conductivity  
of Tl-2201. 
The results shown in (a) and (b) have been obtained 
using the assignment {\it A1}
(the Tl-layer oxygens vibrate at the highest frequency,
the apical oxygens at $\sim 400{\rm\,cm^{-1}}$), 
those shown in (c) and (d) correspond to the reverse assignment ({\it A2}). 
The results shown in (a) and (c) 
have been obtained considering the region A as ``intra-bilayer'', 
those shown in (b) and (d) have been obtained 
considering the region A as ``inter-bilayer''. 
The planar oxygen mode manifests itself 
as a step around $240{\rm\,cm^{-1}}$ 
in (b) and (d). 
The dashed and solid lines correspond to the two values 
of the plasma frequency of region A, $\Omega_{A1}$ and $\Omega_{A2}$, 
$\Omega_{A1}<\Omega_{A2}$,  
respectively, as discussed in the text. 
The dotted and dashed-dotted lines in (d) 
represent results for $\Omega_{A}=1500{\rm\,cm^{-1}}$
and $\Omega_{A}=600{\rm\,cm^{-1}}$, respectively. 
The inset of (b) displays 
the step in the experimental data of Ref.~\onlinecite{Basov1}.
The inset of (c) displays 
the experimental spectra of $(N_{n}-N_{s})/\rho_{s}$
taken from Ref.~\onlinecite{Basov1}. 
The arrow demonstrates the increase of the spectral weight 
of the phonon peak at $390{\rm\,cm^{-1}}$ 
in the superconducting state, 
as discussed in the text. 
}

\figure{                                                                                          
Crystal structure of Tl$_{2}$Ba$_{2}$CuO$_{6}$.                                                   
}                                                                                                 

\figure{
(a) Schematic representation of the model. 
(b) The average electric field $E_{bl}$ 
between the charged planes corresponding to the planar oxygens 
(thick horizontal lines) 
displaced from their equilibrium positions
(thin horizontal lines) 
possesses a contribution $\Delta E_{bl}$ 
due to the charged Y-plane (dashed line), 
$\Delta E_{bl}=[((d_{bl}/2)+a)E_{Y}+((d_{bl}/2)-a)(-E_{Y})]/d_{bl} $. 
Here $E_{Y}$ is the field due to the Y-plane 
in the upper region of the figure. 
A slightly modified version of Fig.~2 from Ref.~\onlinecite{PhonanJLTP}. 
} 


\begin{thebibliography}{La86}
\bibitem{albega}
For Bi-2212,\\ 
$\alpha=-[n_{Ca}e^{*}_{Ca}+(n_{Cu}e^{*}_{Cu}/2)](d_{bl}+d_{int})/ 
         (n_{O(1)}e^{*}_{O}\,d_{bl})$, \\ 
$\beta=-[n_{Bi}e^{*}_{Bi}+n_{O(3)}e^{*}_{O}+n_{Ba}e^{*}_{Ba}+
         n_{O(2)}e^{*}_{O}+(n_{Cu}e^{*}_{Cu}/2)](d_{bl}+d_{int})/ 
         (n_{O(1)}e^{*}_{O}\, d_{int})$, and \\
$\gamma=(d_{bl}+d_{int})/d_{int}$,\\ 
where, e.g., $n_{Ca}$ is the number of Ca ions per unit cell, 
O(1) denotes the planar oxygen, O(3) the BiO-layer oxygen, 
and O(2) the apical oxygen.  
For Tl-2201,\\                                               
$\alpha=-[n_{Ba}e^{*}_{Ba}+n_{O(3)}e^{*}_{O}+n_{Cu}e^{*}_{Cu}]
  (d_{A}+d_{B})/                                    
  (n_{O(2)}e^{*}_{O}\,d_{A})=1.1$, \\               
$\beta=-[n_{Tl}e^{*}_{Tl}+n_{O(1)}e^{*}_{O}](d_{A}+d_{B})/                    
         (n_{O(2)}e^{*}_{O}\, d_{B})=0.9$, and \\                               
$\gamma=(d_{A}+d_{B})/d_{B}=1.8$,\\                                             
where O(1) denotes the TlO-layer oxygen, O(2) the apical oxygen,               
and O(3) the planar oxygen. 
These values correspond to the case 
where the region A is considered as ``intra-bilayer''. 
In the other case (A is considered as ``inter-bilayer'') 
the value of $\alpha$ ($\beta$) 
corresponds to the above value of $\beta$ ($\alpha$) 
and  
$\gamma=(d_{A}+d_{B})/d_{A}=2.2$.
\bibitem{EJ} The formula is equivalent to formula (5) 
of Ref.~\onlinecite{PWA2} or formula (1) of Ref.~\onlinecite{Schutzman1}. 
It differs by a factor of 4 from formula (4) of Ref.~\onlinecite{Chakravarty2}.
\bibitem{background} The values of $\omega_{bl}$ can be simply guessed 
from the data by inserting the frequency 
of the transverse plasma excitation into Eq.~(A.2)
and using an estimate of $\varepsilon_{\infty}$. 
However, the plasma frequency is screened not only 
by the interband processes contributing to $\varepsilon_{\infty}$ 
but also by midinfrared processes. 
The description of the contribution of these processes  
contains a certain degree of arbitrarness. 
In particular, when fitting the infrared data, 
one has to make some assumptions 
concerning the frequency dependendence of this contribution 
and concerning its distribution between the intra- and inter-bilayer 
regions. 
Note that the assumptions made in Ref.~\onlinecite{Phonan}  
are different from those made in Ref.~\onlinecite{Gruninger}.
\bibitem{aboveTc} In Ref.~\onlinecite{Phonan}
we have suggested 
that the onset of the anomalies above $T_{c}$
can be explained by assuming 
that the intra-bilayer plasmon starts to develop already below ${T^{*}>T_{c}}$. 
There is no global phase coherence above $T_{c}$
and consequently, strictly speaking, no Josephson effects. 
It can be speculated, however, that above $T_{c}$ there is  
either a certain degree of phase coherence 
between the closely spaced copper oxygen planes
or at least some kind of pairing developed 
making a (semi)coherent intra-bilayer hopping possible.
\bibitem{scaleofHj} This scale corresponds to the frequency range 
where the single particle tunneling is completely blocked.
\bibitem{cutoffs} In reality the regions of intraband and interband 
absorption overlap. Considering this would make the discussion 
more complicated, while it would not change our main conclusions.
\bibitem{countereffects} The contribution to the $c$-axis kinetic energy 
due to the single particle tunneling may increase below $T_{c}$; 
that is what we mean by ``countereffects''.
\bibitem{factor4} For a given plasma frequency of the condensate 
(or, alternatively, for a given penetration depth) the sum rule-based-approach 
of Refs.~\onlinecite{Chakravarty2,Chakravarty3} (Eq.~(8)) 
yields 
a change of the $c$-axis kinetic energy 
by a factor of 4 larger than Eq.~(3) 
if the changes of the regular part of $\sigma_{1}(\omega)$ 
between the normal and the superconducting state are negligible. \\
For underdoped La$_{2-x}$Sr$_{x}$CuO$_{4}$ this discrepancy
is partially reduced because of the changes in the background.
\bibitem{Y123Tc53} This is particularly clear in case                             
of underdoped Y123 with $T_{c}=53{\rm\,K}$                                        
(see Fig.~1 (a)).  Both the additional peak and the anomaly                       
of the $320{\rm\,cm^{-1}}$ phonon mode can be fitted                              
using only three essential parameters:                                            
$\omega_{bl}$, $S_{P}$ (oscillator strength of the phonon)                        
and $\omega_{P}$ (frequency of the phonon).
\bibitem{Anomaly} It is suprising that the Tl-layer-oxygen mode
exhibits the anomalies rather than the apical-oxygen mode 
(the apical-oxygen plane forms a boundary between the regions 
A and B in the same way as the plane of the planar oxygens 
in Y123 forms a boundary 
between the intra-bilayer and the inter-bilayer regions). 
The reason is that the geometry is different. 
In Y123, $d_{int}>>d_{bl}$, 
here $d_{A}\approx d_{B}$.
\bibitem{absvalues}                                                                      
The values of the model conductivity                                                     
in the frequency region around $800{\rm\,cm^{-1}}$                                       
are by a factor of $\sim 2$ higher                                                       
than the value of $\sim 10{\,\Omega^{-1}{\rm\,cm^{-1}}}$                                 
reported in Ref.~\onlinecite{Basov1}.                                                    
This is not a serious discrepancy                                                        
considering the oversimplified ansatz for $\sigma_{A}$ and $\sigma_{B}$.
\end{thebibliography}
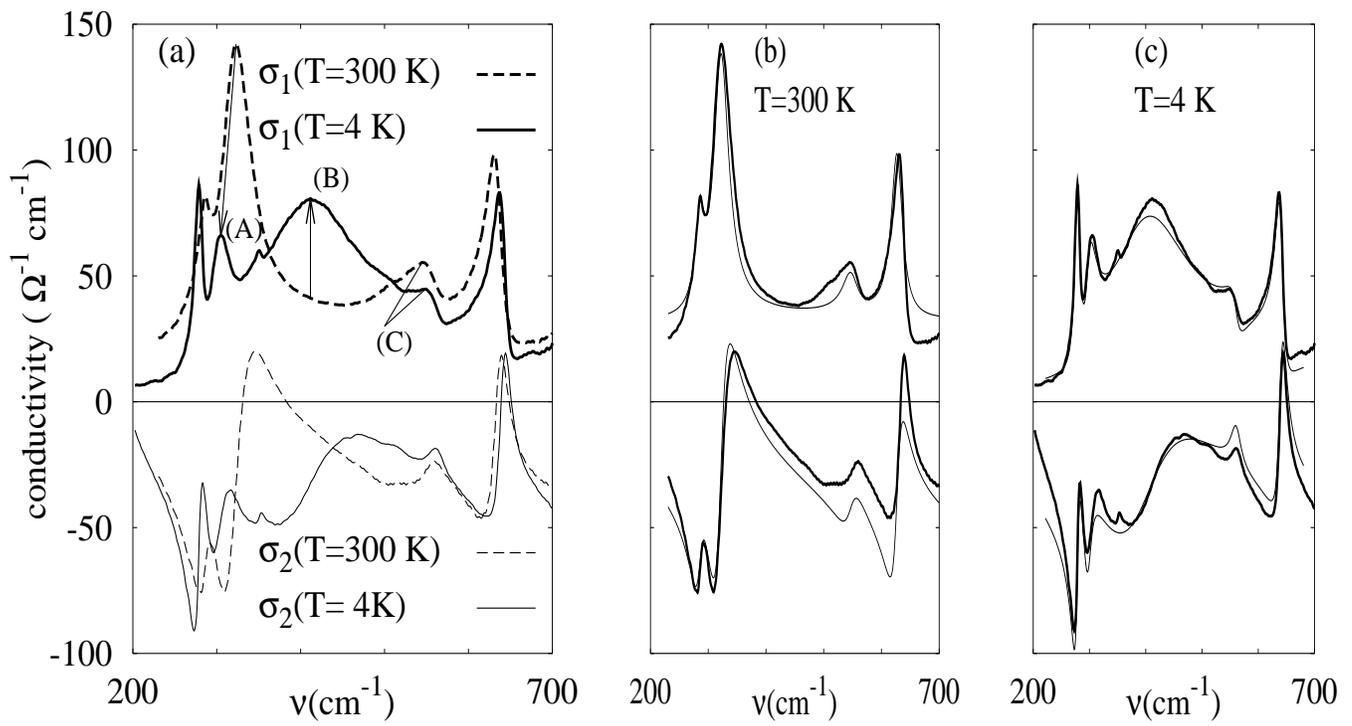
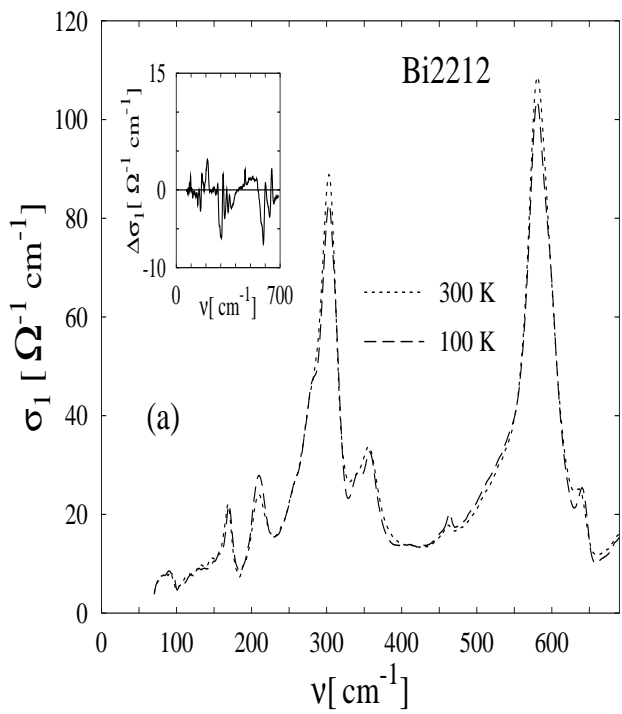
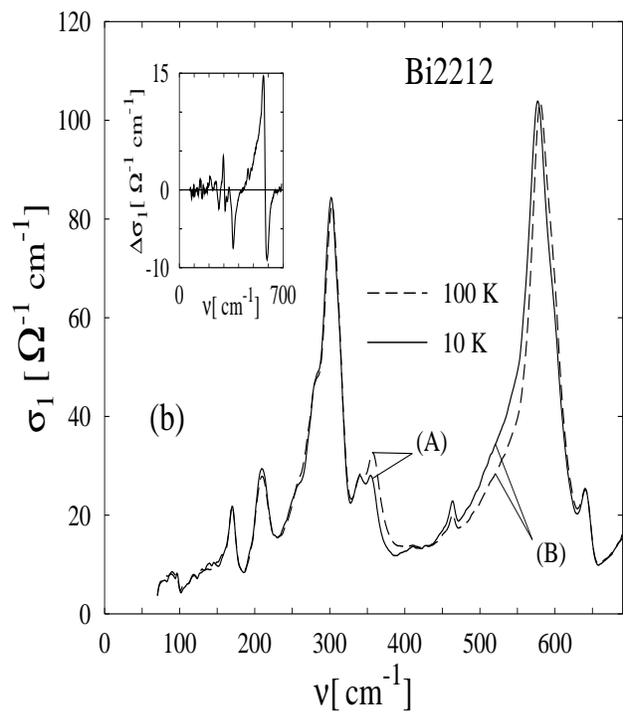
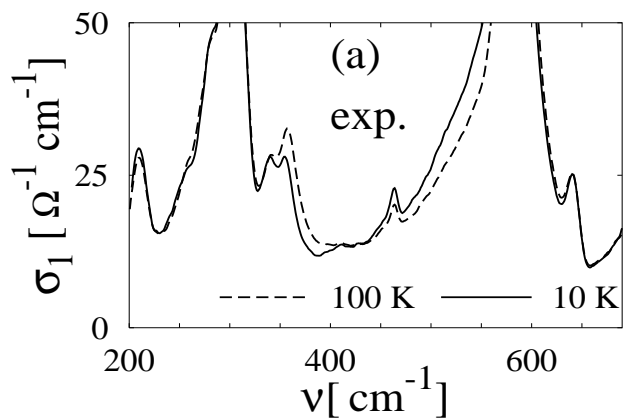
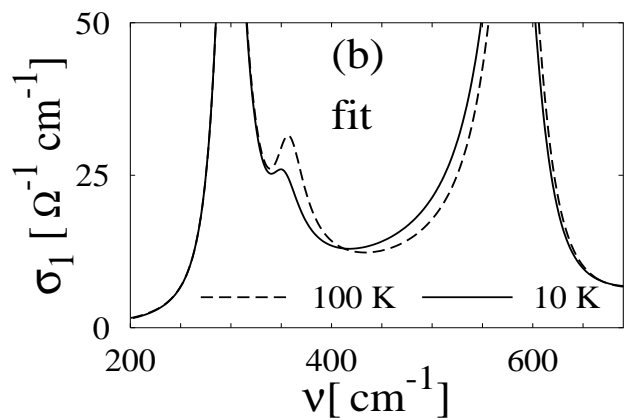
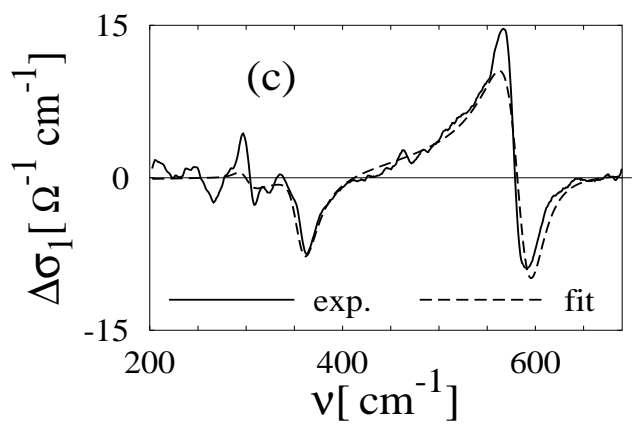
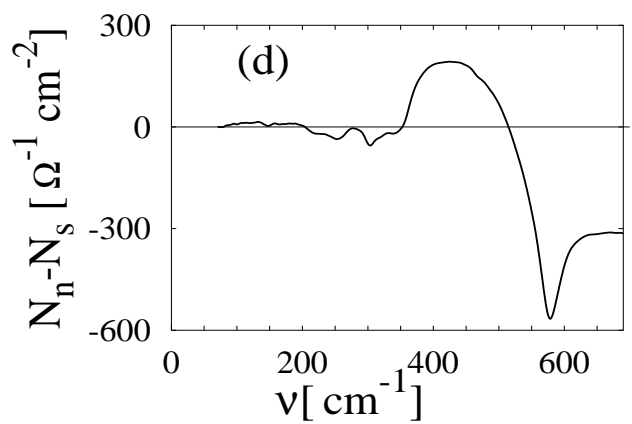
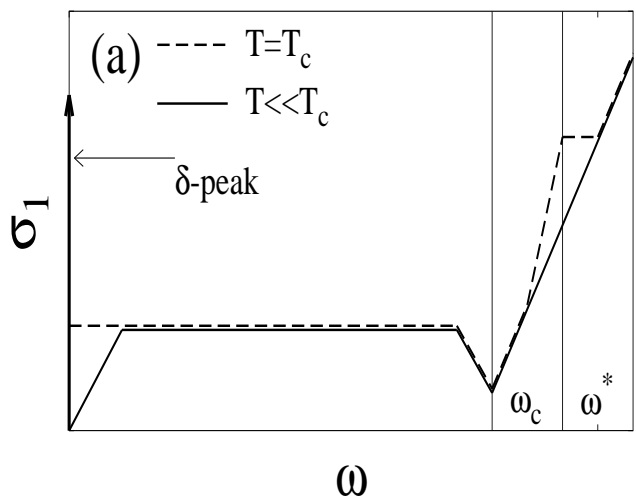
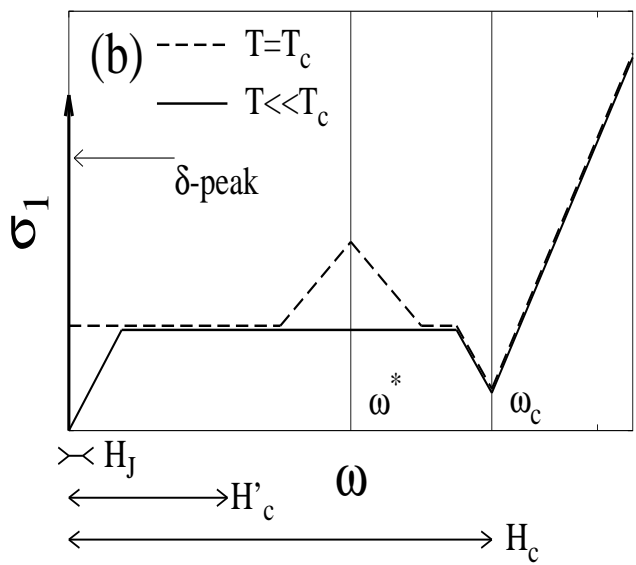
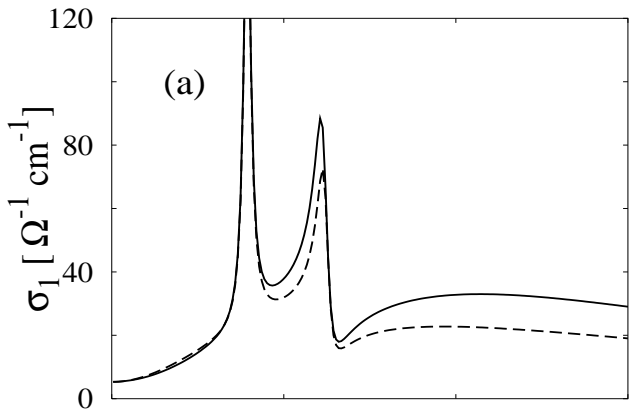
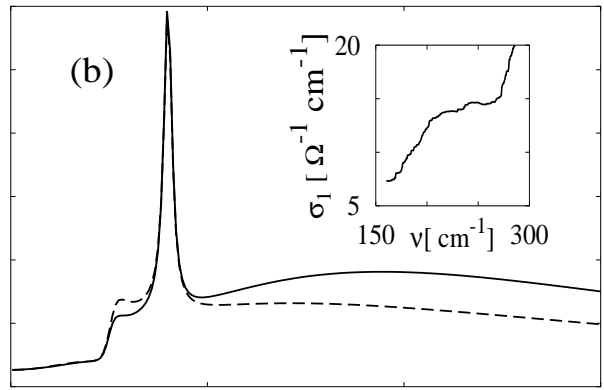
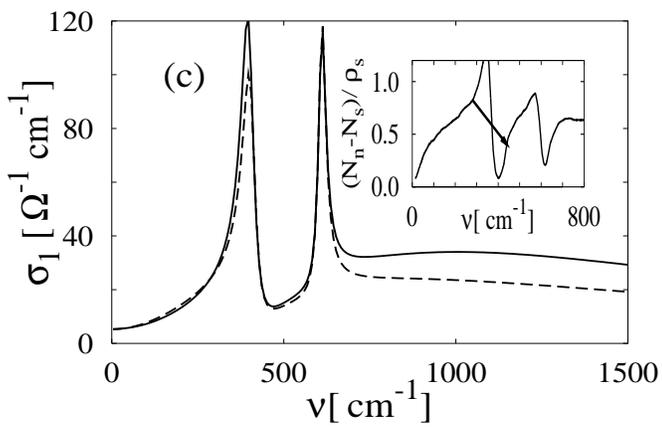
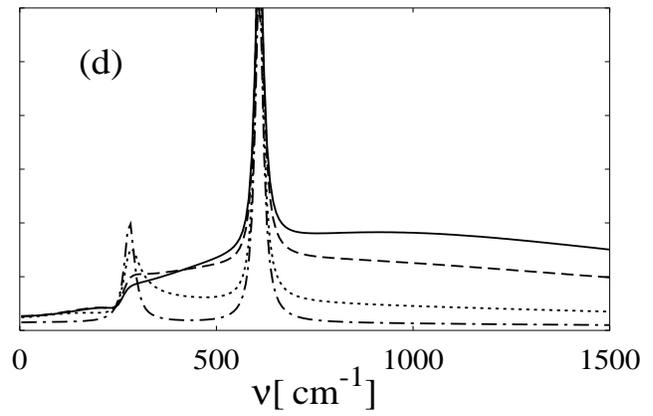
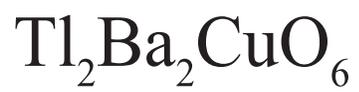
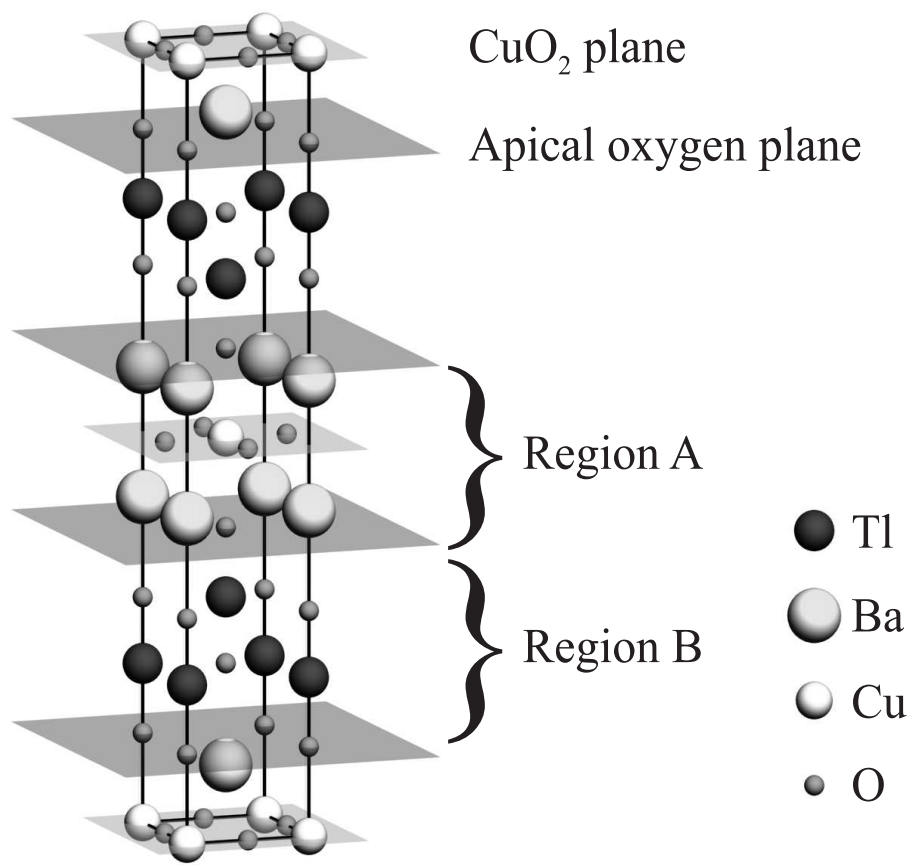
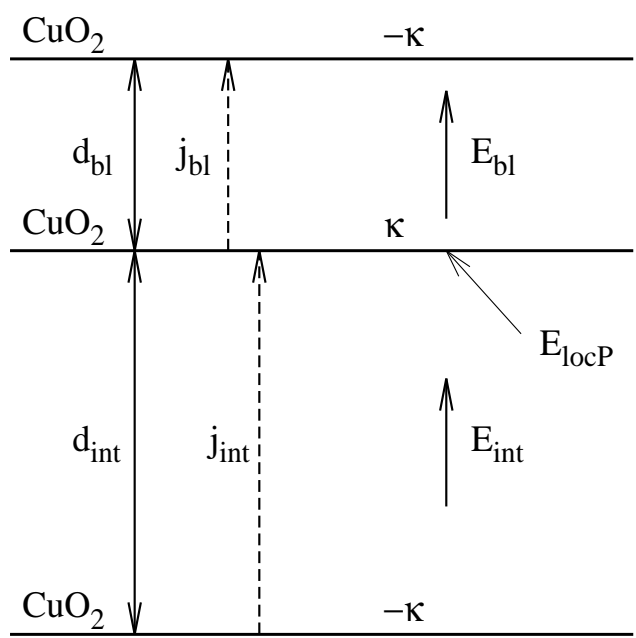
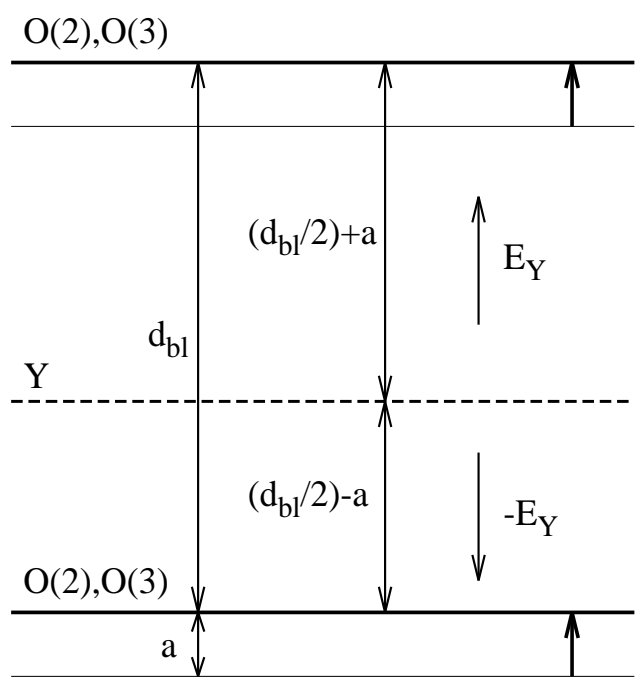
\end{document}